\documentclass[referee]{aa} 
\usepackage{graphicx}
\usepackage{txfonts}
\begin{document}
   \title{Molecular outflows towards O-type young stellar objects\thanks{Based on observations carried out with the IRAM~30-m telescope on Pico Veleta (Granada, Spain). IRAM is supported by INSU/CNRS (France), MPG (Germany), and IGN (Spain).}}


   \author{A. L\'opez-Sepulcre
          \inst{1}
          \and
	  C. Codella
	  \inst{1}
	  \and
	  R. Cesaroni
	  \inst{1}
	  \and
	  N. Marcelino
	  \inst{2}
          \and
          C.M. Walmsley
          \inst{1}}

\institute{INAF, Osservatorio Astrofisico di Arcetri, Largo E. Fermi 5, 50125 Firenze,
Italy\\
\email{sepulcre@arcetri.astro.it, codella@arcetri.astro.it, cesa@arcetri.astro.it, walmsley@arcetri.astro.it}
\and
Laboratorio de Astrof\'isica Molecular, CAB-CSIC/INTA, Ctra de Torrej\'on a Ajalvir km 4, 28850 Torrej\'on de Ardoz, Madrid, Spain\\
\email{nuria@damir.iem.csic.es}}

\date{Received date; accepted date}

\abstract{The formation of massive stars is not well-understood and requires detailed observational studies in order to discriminate between the different proposed star formation models.} {We have searched for massive molecular outflows in a sample of high-mass star forming regions, and we have characterised both the outflow properties and those of their associated molecular clumps. With a sample composed largely of more luminous objects than previous ones, this work complements analogous surveys performed by other authors by adding the missing highest luminosity sources.} {The sample under study has been selected so as to favour the earliest evolutionary phases of star formation, and is composed of very luminous objects ($L_\mathrm{bol} > 2\times 10^{4}$~L$_{\sun}$ and up to $\sim$10$^{6}$~L$_{\sun}$), possibly containing O-type stars. Each source has been mapped in $^{13}$CO(2$-$1) (an outflow tracer) and C$^{18}$O(2$-$1) (an ambient gas tracer) with the IRAM-30m telescope on Pico Veleta (Spain).} {The whole sample shows high-velocity wings in the $^{13}$CO(2$-$1) spectra, indicative of outflowing motions. In addition, we have obtained outflow maps in 9 of our 11 sources, which display well-defined blue and/or red lobes. For these sources, the outflow parameters have been derived from the line wing $^{13}$CO(2$-$1) emission. An estimate of the clump masses from the C$^{18}$O(2$-$1) emission is also provided and found to be comparable to the virial masses. From a comparison between our results and those found by other authors at lower masses, it is clear that the outflow mechanical force increases with the bolometric luminosity of the clump and with the ionising photon rate of the associated H{\sc ii} regions, indicating that high-mass stars drive more powerful outflows. A tight correlation between outflow mass and clump mass is also found.} {Molecular outflows are found to be as common in massive star forming regions as in low-mass star forming regions. This, added to the detection of a few tentative large-scale rotating structures suggests that high-mass stars may generally form via accretion, as low-mass stars.}

\keywords{Stars: formation -- ISM: clouds -- ISM: jets and outflows -- ISM: molecules}

\maketitle
%

\def\HII{H{\sc ii}}
\def\WAT{H$_2$O}
\def\MET{CH$_3$OH}
\def\AMM{NH$_3$}
\def\MCN{CH$_3$CN}
\def\kms{km~s$^{-1}\,$}
\def\13co{$^{13}$CO(2$-$1)}
\def\c18o{C$^{18}$O(2$-$1)}
\def\co{$^{12}$CO}

\section{Introduction}

Massive stars have a high impact on their environments and play a major role in the evolution of the Galaxy. However, their formation mechanism is far from being well understood. There is a main theoretical problem that arises when studying massive star formation: unlike low-mass stars, stars with masses greater than $\sim$8~M$_{\sun}$ reach the zero-age main sequence (ZAMS) while they are still accreting material from the surrounding parental molecular cloud (Palla \& Stahler \cite{palla93}). The radiation pressure they develop at this point is expected to exceed the gravitational pressure of the collapsing material, thus halting the accretion process and preventing the star from gaining more mass. This leads to the conclusion that stars more massive than about 8~M$_{\sun}$ cannot form, which is in clear disagreement with observations. The various models which have been proposed to solve this problem can be grouped into two main scenarios: \emph{accretion} through massive circumstellar disks and/or sufficiently high accretion rates (e.g. Yorke \& Sonnhalter \cite{yorke02}, Tan \& McKee \cite{tan02}, Keto et al. \cite{keto03}), and \emph{coalescence} of lower-mass stars in very dense clusters (Bonnell, Bate \& Zinnecker \cite{bonnell98}, Bonnell \& Bate \cite{bonnell02}).

Observing the early phases of the  evolution of massive protostars is rendered more difficult by the  high extinction, the relatively large distance (several kiloparsec) and by the considerable confusion. Massive stars do not form alone but in a cluster and the phenomena associated with the formation, say, of an O-star, have to be disentangled from those associated with the formation of lower mass stars. Reviews of different aspects of the problem are given by Beuther et al. (\cite{beut07}), Hoare et al. (\cite{hoare07}), McKee \& Ostriker (\cite{mckee07}), and Zinnecker \& Yorke (\cite{zin07}). One general conclusion is that feedback due to outflows and ionising radiation from young massive stars may drastically influence the formation of still younger stars.  Outflows and water or methanol masers are often moreover the most obvious signs of a young massive protostar. It is therefore of considerable interest to determine the physical properties of the outflows detected in high-mass star forming regions (SFRs).

Outflows in high-mass SFRs have had considerable study (see e.g. Arce et al. \cite{arce07}). Two interesting examples are the flows associated with G24.78+0.08 and G31.41+0.31. These are two high-luminosity sources showing signposts of high-mass star formation, such as water masers and ultracompact (UC) \HII\  regions, which have been observed in a variety of molecular tracers, both with single-dish telescopes and interferometers (e.g. Cesaroni et al. \cite{cesa94}, Olmi et al. \cite{olmi96}, Furuya et al. \cite{furuya02}, Beltr\'an et al. \cite{beltran04}). The observations reveal disk-outflow systems, and suggest that high-mass stars in these two regions form via accretion inside a pc-scale molecular clump. It is clear from the results of these particular studies that only the combination of low- and high-angular resolution observations can provide us with the necessary information on the structure and kinematics of high-mass SFRs. Despite the great importance of these findings, the formation scenario depicted for G24 and G31 does not necessarily have to represent the general case. Therefore, more examples like these need to be found and investigated in detail in order to gradually build up a statistically significant collection of well-known massive SFRs, and thus understand better the process of high-mass star formation.

Outflow studies to date have  mostly been carried out using $^{12}$CO or SiO. Using SiO is problematic because it seems likely that the SiO abundance varies widely between outflows and perhaps as a function of velocity within a single outflow. Using $^{12}$CO is also difficult because of the effects of optical depth  as illustrated by the work of Choi et al. (\cite{choi93}), who found optical depths of the order of 10 in the CO(3$-$2) line towards a couple of flows in high mass star forming regions. They also found that the optical depth dropped off sharply in velocity channels well separated from the ambient value.  High optical depth in $^{12}$CO has also to be contended with in the lower $J$ transitions and has the consequence that outflow mass estimates based on the assumption of low optical depth in  $^{12}$CO  are underestimates. Partly for this reason, we  have decided to estimate outflow mass and momentum in this article using the \13co\ line and correcting  for possible finite optical depth in this transition using the wings of \c18o. While we expect this technique will yield a more accurate outflow mass estimate than that from $^{12}$CO, one misses in this way the high velocity CO wings (full width at zero power above 100~km~s$^{-1}$ in some of the Choi et al. sample) and thus we underestimate the mechanical energy output.

As a starting point for this study, we have conducted a small survey towards 11 high-mass star forming regions. This allows us to derive physical properties of a reasonable number of objects and, perhaps most importantly, to select the best candidates for follow-up observations at higher spatial resolutions. A search for massive molecular outflows at millimetric wavelengths is well-suited for a single-dish project, since high-mass outflows are extended structures with sizes of $\sim$1~pc (e.g. Shepherd \& Churchwell \cite{shepherd96}, Beuther et al. \cite{beut02a}), i.e. a few half-power beam widths (HPBWs) at the typical distances of these sources (1-10~kpc). Moreover, the detection of a bipolar molecular outflow in a high-mass SFR already suggests the presence of an accretion disk and encourages further investigation at higher-angular resolution to search for direct signatures of rotating disks/toroids and more collimated jets closer to the central object. 

With this idea in mind, we have carried out a \13co\ and \c18o\ survey with the IRAM~30-m telescope (Granada, Spain) towards a sample of selected high-mass young stellar objects (YSOs) in their earliest evolutionary stages, to search for massive molecular outflows. We describe the sample and the observations in Sect. 2. The results of our single-dish data are presented in Sect. 3. In Sect. 4 we derive the physical parameters of the outflows and the molecular clumps hosting them, in Sect. 5 we discuss the general results and comment on some particular sources, and finally, Sect. 6 summarises our results and conclusions.

\section{Observations and data reduction}

\subsection{The sample}

The sample observed in the present study is composed of 11 high-mass SFRs chosen from the \WAT\ maser surveys by Hofner \& Churchwell (1996) and Forster \& Caswell (1999), and from the \MET\ maser survey by Walsh et al. (1997). All the sources in our sample have bolometric luminosities greater than $2\times 10^{4}$~L$_{\sun}$ (i.e. possibly with O-type stars) and present compact or no free-free emission, i.e. they are associated with either UC \HII\ or no detected \HII\ regions.

In addition, they are associated with at least one of the following: (i) \WAT\ masers, (ii) \MET\ masers, (iii) compact (sub)mm continuum emission. All these criteria ensure the selection of high-mass YSOs in their earliest evolutionary phases.

\begin{table*}[!hbt]
\caption{The sample: coordinates used for the observations, distance, bolometric luminosity, ionising photon rate, methanol maser detection and outflow detection}
\label{sample}
\vspace{1mm}
\begin{center}
\begin{tabular}{cccccccccc}
\hline
Source & $\alpha$ (2000) & $\delta$ (2000) & $d$ & $d_\mathrm{gal}$$^\mathrm{\dag}$ & $L_{\mathrm{bol}}$$^\mathrm{\ddag}$ & $N_{\mathrm{Ly}}$ & CH$_3$OH masers?$^\mathrm{\star}$ & H$_2$O masers? & Outflow?\\
 & (h m s) & (\degr~\arcmin~\arcsec) & (kpc) & (kpc) & 10$^4$~L$_{\sun}$ & (10$^{47}$ s$^{-1}$) & (Y/N) & (Y/N) & (Y/N)\\
\hline
G10.47$+$0.03 & 18 08 38.22 & $-$19 51 49.7 & 10.8$^\mathrm{a}$ & 2.9 & 140$^\mathrm{f}$ & 28$^\mathrm{k}$ & Y & Y$^\mathrm{b}$ & Y\\
G10.62$-$0.38 & 18 10 28.70 & $-$19 55 49.7 & 6.5$^\mathrm{b}$ & 2.4 & 110$^\mathrm{g}$ & 91$^\mathrm{l}$ & Y & Y$^\mathrm{b}$ & Y\\
G16.59$-$0.06 & 18 21 09.16 & $-$14 31 48.8 & 4.7$^\mathrm{c}$ & 4.2 & 2.3$^\mathrm{g}$ & 0.26$^\mathrm{m}$ & Y & Y$^\mathrm{d}$ & Y\\
G19.61$-$0.23 & 18 27 38.15 & $-$11 56 38.5 & 12$^\mathrm{a}$ & 4.9 & 170$^\mathrm{g}$ & 49$^\mathrm{l}$ & N & Y$^\mathrm{b}$ & Y\\
G23.44$-$0.18 & 18 34 39.25 & $-$08 31 38.8 & 7.8$^\mathrm{d}$ & 3.4 & 57$^\mathrm{g}$ & $<$0.02$^\mathrm{n}$ & N & Y$^\mathrm{d}$ & ?\\
G28.87$+$0.06 & 18 43 46.24 & $-$03 35 30.4 & 8.5$^\mathrm{d}$ & 4.2 & 29$^\mathrm{g}$ & 0.13$^\mathrm{m}$ & Y? & Y$^\mathrm{d}$ & Y\\
G29.96$-$0.02 & 18 46 03.96 & $-$02 39 21.5 & 9$^\mathrm{e}$ & 4.5 & 190$^\mathrm{g}$ & 31$^\mathrm{k}$ & Y & Y$^\mathrm{b}$ & ?\\
G35.20$-$0.74 & 18 58 12.98 & $+$01 40 37.6 & 2.3$^\mathrm{d}$ & 6.8 & 2.1$^\mathrm{h}$ & 0.02$^\mathrm{n}$ & Y? & Y$^\mathrm{d}$ & Y\\
G43.89$-$0.78 & 19 14 26.16 & $+$09 22 34.0 & 4.2$^\mathrm{b}$ & 6.2 & 4.0$^\mathrm{f}$ & 9.1$^\mathrm{l}$ & Y? & Y$^\mathrm{b}$ & Y\\
G48.61$+$0.02 & 19 20 31.19 & $+$13 55 24.9 & 11.8$^\mathrm{d}$ & 8.9 & 130$^\mathrm{i}$ & 4.4$^\mathrm{n}$ & -- & Y$^\mathrm{d}$ & Y\\
G75.78$+$0.34 & 20 21 44.10 & $+$37 26 39.5 & 4.1$^\mathrm{b}$ & 8.5 & 24$^\mathrm{j}$ & 3.9$^\mathrm{l}$ & Y? & Y$^\mathrm{b}$ & Y\\
\hline
\end{tabular}
\end{center}
\begin{itemize}
\item [] \footnotesize{$^\mathrm{\dag}$ Galactocentric distance}
\item [] \footnotesize{$^\mathrm{\ddag}$ Luminosities and ionising photon rates corrected for revised distances}
\item [] \footnotesize{$^\mathrm{\star}$ Pestalozzi et al. (\cite{pestalozzi05}). Sources with ''?'' have only single-dish observations; ''--'' denotes no observations towards that source.}
\item [] \footnotesize{References: $^\mathrm{a}$ Pandian et al. (\cite{pandian08}),
$^\mathrm{b}$ Hofner \& Churchwell (\cite{hofner96}),
$^\mathrm{c}$ Codella et al. (\cite{codella97}),
$^\mathrm{d}$ Forster \& Caswell (\cite{forster99}),
$^\mathrm{e}$ Hanson et al. (\cite{hanson02}), $^\mathrm{f}$ Hatchell et al. (\cite{hatchell00}), $^\mathrm{g}$ Walsh et al. (\cite{walsh97}), $^\mathrm{h}$ Fuller et al. (\cite{fuller01}), $^\mathrm{i}$ Kurtz et al. (\cite{kurtz94}), $^\mathrm{j}$ Klaassen \& Wilson (\cite{klaassen07}), $^\mathrm{k} $Cesaroni et al. (\cite{cesa94}), $^\mathrm{l}$ Wood \& Churchwell (\cite{wood89}), $^\mathrm{m}$  Cesaroni (priv. comm.), $^\mathrm{n}$ Codella (priv. comm.)}
\end{itemize}
\end{table*}

In Table \ref{sample} we present the list of the observed sources, their coordinates, distances and bolometric luminosities. The last three columns show, respectively, the rate of ionising photons from the associated ultracompact (UC) \HII\  regions, the methanol maser detection as reported by Pestalozzi et al. (2005), and the molecular outflow detection according to our observations (see Sect.~3.1). Note that all the sources are associated with UC \HII\ regions but G23.44, where no ionised emission has been detected so far. For this source, an upper limit for N$_\mathrm{Ly}$ is given, determined from 3$\sigma$. Both $L_{\mathrm{bol}}$ and $N_{\mathrm{Ly}}$ have been corrected for revised distances, based on the references in Table \ref{sample}.

\subsection{IRAM 30-m observations}

The IRAM 30-m telescope at Pico Veleta (near Granada, Spain) was used to map the 11 sources of our sample in the \13co and \c18o lines. We have chosen the less abundant $^{13}$CO isotopologue instead of the more commonly used outflow tracer $^{12}$CO in order to decrease the effects due to confusion with other unrelated clouds along the line of sight, which is likely, given the galactic distribution of the objects (see also the discussion in Sect. 5.1). The observations were carried out on September 23 and 24, 2006. The average $T_{\mathrm{sys}}$ during the observations was $\sim$1000~K for the \13co and $\sim$1400~K for the \c18o line. The pointing was checked about every hour by observing nearby planets and it was found to be accurate to within 4\arcsec. 

The maps have sizes of 4\arcmin$\times 4$\arcmin\ and were obtained with the 3$\times$3 pixel receiver array HERA in the on-the-fly mode, and the autocorrelator VESPA. At the frequency of the lines, i.e. 220~GHz, the resolution of the 30-m telescope is about 11$''$. The dump time was 2~s, and the sampling interval 4$''$. All the maps except one were scanned twice, once along the R.A. direction and once along the Dec. direction, to avoid scanning effects on the resulting images. The map of G19.61$-$0.23 was scanned only along the R.A. direction due to lack of time. The velocity resolution is $\sim$0.1~km~s$^{-1}$ (80~kHz), and the bandwidth $\sim$110~\kms (80~MHz).

The data were reduced and analysed with the programs CLASS and GRAPHIC of the GILDAS software package developed by the IRAM and the Observatoire de Grenoble. Both the \13co and the \c18o spectra have been smoothed to a resolution of 0.5~km~s$^{-1}$, in order to improve the signal to noise ratio. After smoothing, the average rms level for a single spectrum is $\sim$2.5~K, and the average rms of the outflow maps is $\sim$0.3~K.

\section{Results}

\subsection{Molecular outflows}

\begin{figure*}[thb]
\centering
\includegraphics[width=12cm]{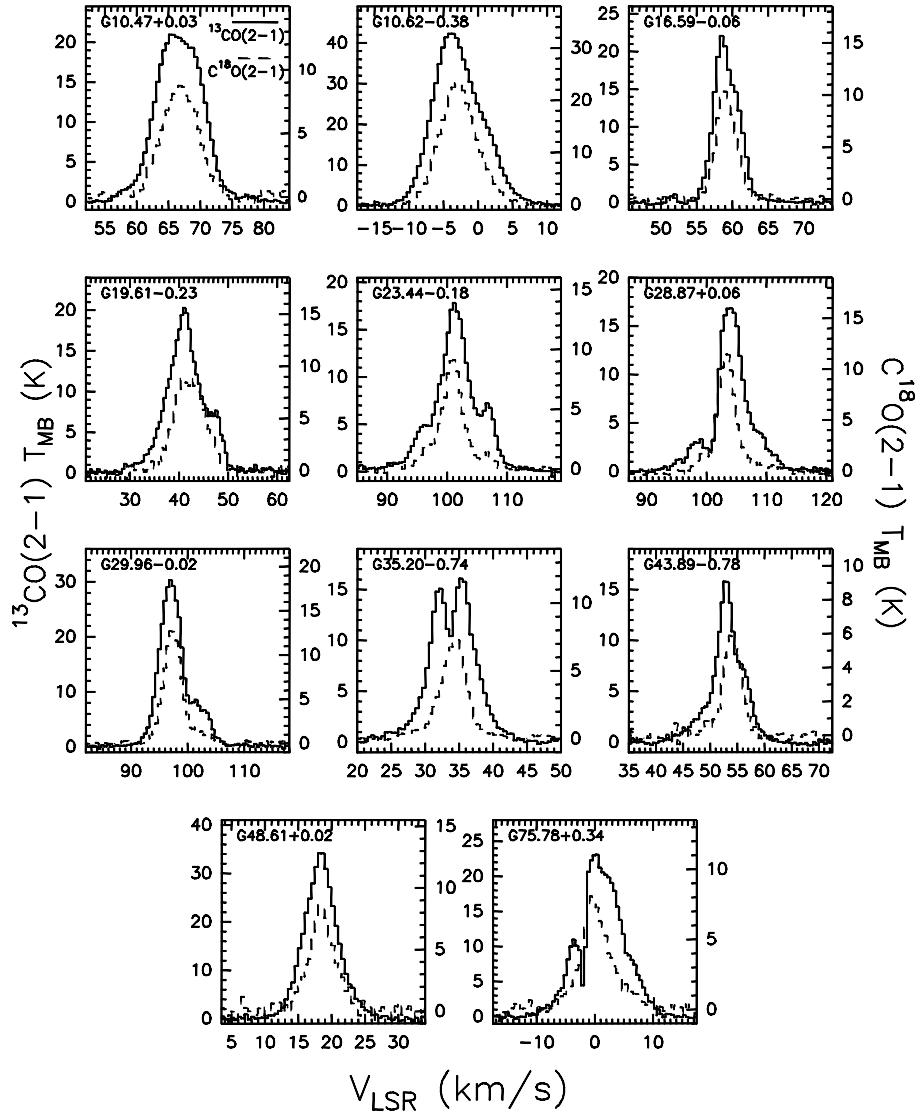}
\caption{\13co (solid) and \c18o (dashed) spectra of each source, obtained by integrating the emission inside the 50\% contour level of the \c18o maps. In each box the scale to the left corresponds to the \13co\ main beam brightness temperature, the one to the right that of \c18o.}
\label{spec2}
\end{figure*}

\begin{figure*}[!bht]
\centering
\includegraphics[width=18cm]{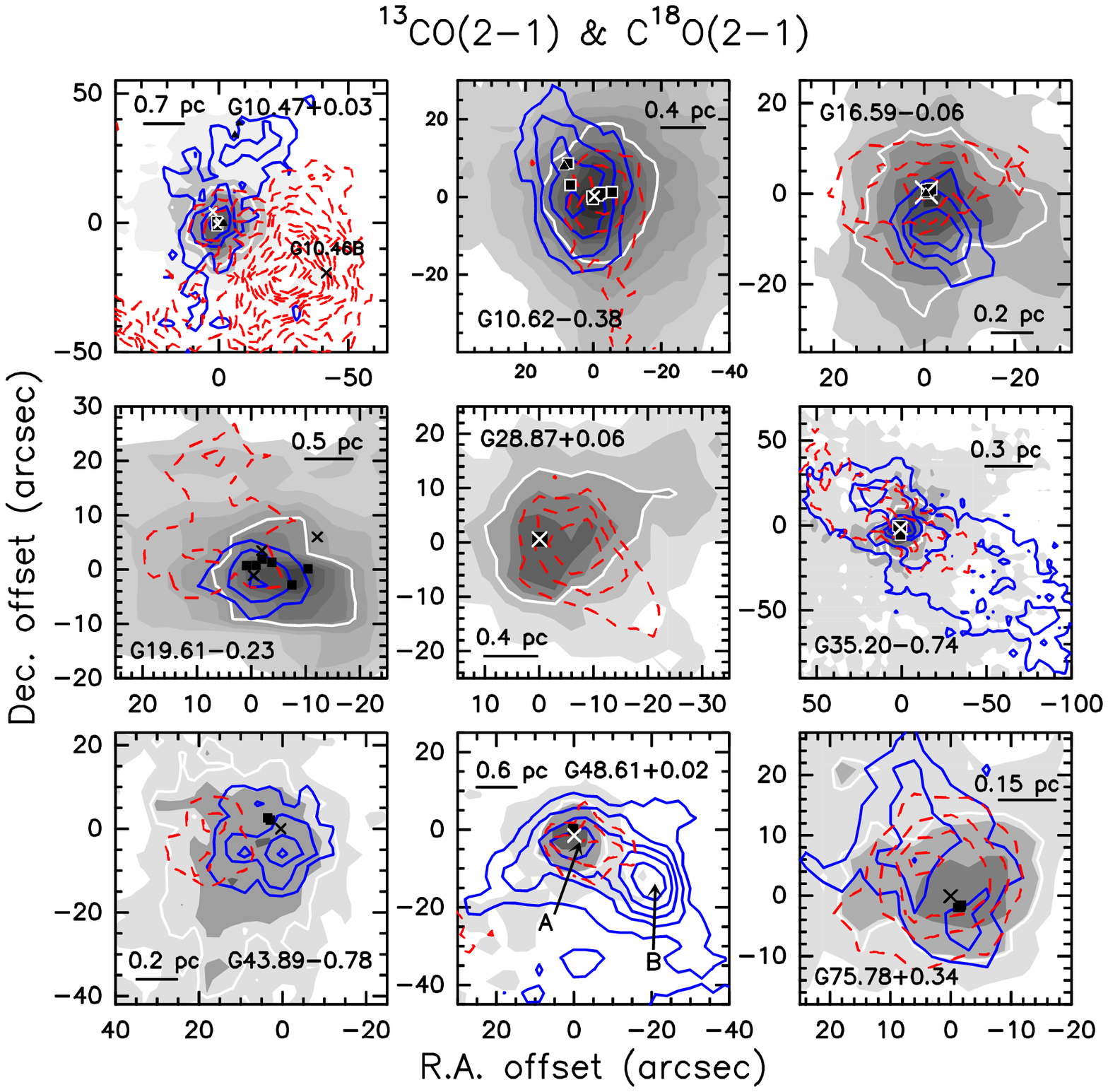}
\caption{Molecular outflows: the solid lines represent the blue wings and the dashed lines the red wings of the \13co emission, with contours starting from 5$\sigma$ and increasing by steps of 2$\sigma$ (G43), 3$\sigma$ (G10.47, G16, G19, G28, G48, G75) and 5$\sigma$ (G10.62, G35). The rms values of the blue emission are 0.2 (G10.47, G35), 0.3 (G10.62, G16, G19) and 0.4 (G43, G48, G75)~K. The rms values of the red emission are 0.2 (G10.47, G16, G35), 0.3 (G10.62, G19, G28), 0.4 (G48) and 0.5 (G43, G75)~K. The blue-wing and red-wing velocity ranges used for integration are shown in Table \ref{outcal}. The grey scale depicts the corresponding \c18o emission from 5$\sigma$ by steps of 5$\sigma$. The rms values of the \c18o emission are 0.2 (G10.47, G19), 0.3 (G16, G35), 0.4 (G10.62, G28, G43) and 0.5 (G48, G75) K. The \c18o half-power contour is drawn in white. Triangles and squares represent CH$_3$OH and H$_2$O maser spots, respectively (from Walsh et al. \cite{walsh98}, Hofner \& Churchwell, \cite{hofner96}, and Forster \& Caswell, \cite{forster99}), and crosses mark the position of UC \HII\ regions (from the authors indicated in column 6 of Table \ref{sample}). The axes show offsets in arcsec from the absolute coordinates given in Table \ref{sample}. We note that the blue emission peak indicated with an arrow as B in the map of G48 corresponds to a different clump with a slightly bluer systemic velocity than the target clump A.
}
\label{outflows}
\end{figure*}

In Fig. \ref{spec2} we present the \13co (solid) and \c18o (dashed) spectra for each observed source. These have been obtained by averaging the emission inside the half power contour of the C$^{18}$O integrated maps (white solid contour in Fig. \ref{outflows}). A comparison between the \13co and the optically thinner \c18o lines allows us to define the systemic velocity, $V_{\mathrm{LSR}}$, of the molecular clumps, identify the \element[][13]{CO} line wings and discriminate between self-absorption and multiple components in the \13co line profiles.

All the \13co spectra display high-velocity wings, typical of outflowing motions.  Outflow maps are shown in Fig. \ref{outflows}, where the blue- and the red-wing emission contours (solid and dashed, respectively) of the \13co line are overlayed on the \c18o integrated map (grey scale). \WAT\ and \MET\ maser spots are also indicated on the maps as filled squares and triangles, respectively, and the positions of the UC \HII\ regions are marked as crosses.

Out of the 11 sources observed, 9 show red- and/or blue-shifted lobes. The remaining two sources (G23.44 and G29.96) both show high velocity \13co\ wings and hence presumably have outflows also (in the case of G29.96, see discussion of Beuther et al. \cite{beut07g29}) but they are sufficiently confused with our angular resolution that we do not discuss them further here. Higher angular resolution maybe in a different tracer is needed to resolve any possible outflow in these regions. The red emission in G10.47 is dominated by a secondary nearby clump towards the south-west, marked as G10.46B in Fig. \ref{outflows} because of its proximity to the UC \HII\ region bearing the same name (Cesaroni et al. \cite{cesa94}; see Appendix). As a consequence, a well-defined isolated red lobe does not appear on the map at the resolution of the observations. However, the elongated blue lobe, together with the wide wings of the \13co spectrum favour the presence of a molecular outflow in this region.

\subsection{Molecular clumps}

The good spatial overlap between the C$^{18}$O bulk emission and the \WAT\ and \MET\ maser spots (Fig. \ref{outflows}) confirms that the C$^{18}$O molecule is efficiently tracing the high-density clump hosting the massive YSOs. The white solid line in each map represents the 50\% contour level of the \c18o integrated emission and gives a measure of the size of the molecular clump. In columns 2 and 3 of Table \ref{mass} the deconvolved linear diameter of the C$^{18}$O clumps, $D$, and the optical depth of the \c18o emission, $\tau_{\mathrm{C^{18}O}}$, are presented. The latter has been estimated from the \13co to \c18o intensity ratio measured at the systemic velocity, using the spectra in Fig. \ref{spec2} and assuming the same excitation temperature, $T_{\mathrm{ex}}$, for both species. A relative abundance of 0.14 has been assumed between the C$^{18}$O and $^{13}$CO isotopologues (Wilson \& Rood \cite{wilson94}).

\begin{table*}[!tbh]
\caption{Deconvolved clump diameter and mass estimates}
\label{mass}
\vspace{1mm}
\centering
\begin{tabular}{ccccccccc}
\hline
Source & $D$ & $\tau_{\mathrm{C^{18}O}}$$^{\mathrm{a}}$ & $M_{\mathrm{clump}}$$^\mathrm{b}$ & $M_{\mathrm{vir}}$$^{\mathrm{c}}$ & T$_\mathrm{dust}$ & $M_{\mathrm{dust}}$$^\mathrm{d}$ & $V_{\mathrm{rot}}$$^\mathrm{e}$ & $M_{\mathrm{dyn}}$[$^\mathrm{f}$]\\
 &  (pc) & & ($M_{\mathrm{\odot}}$) & ($M_{\sun}$) & ($M_{\sun}$) & ($M_{\sun}$) & (km~s$^{-1}$) & ($M_{\sun}$)\\
 \hline
G10.47 & 1.25 & 0.57 & 1460$-$3200 & 5680 & 20$^1$ & 47900$^1$ & 0.57 & 44\\
G10.62 & 0.98 & 0.82 & 1920$-$4110 & 4770 & 20$^1$ & 20200$^1$ & 0.61 & 27\\
G16.59 & 0.77 & 0.69 & 390$-$870 & 1210 & 20$^1$ & 1130$^1$ & --- & ---\\
G19.61 & 1.19 & 0.60 & 2450$-$5160 & 5320 & 20$^1$ & 33000$^1$ & --- & ---\\
G23.44 & 1.79 & 0.76 & 2390$-$5120 & 2850 & 20$^1$ & 4160$^1$ & --- & ---\\
G28.87 & 0.96 & 1.40 & 1260$-$2790 & 1350 & 33$^2$ & 3420$^2$ & --- & ---\\
G29.96 & 1.34 & 0.56 & 2030$-$4460 & 2030 & 20$^1$ & 12900$^1$ & --- & ---\\
G35.20 & 0.32 & 0.83 & 110$-$250 & 460 & 20$^3$ & 1500$^3$ & 0.66 & 14\\
G43.89 & 0.91 & 0.85 & 570$-$1200 & 2230 & 10$^4$ & 6610$^4$ & 0.86 & 88\\
G48.61 & 0.83 & 0.27 & 1200$-$2620 & 1280 & 29$^5$ & 3790$^5$ & --- & ---\\
G75.78 & 0.54 & 0.41 & 450$-$1000 & 1390 & --- & --- & 0.68 & 23\\
 \hline
\end{tabular}
\begin{list}{}{}
\item[$^{\mathrm{a}}$] \footnotesize{Optical depth of the \c18o\ emission, estimated from the \13co to \c18o line intensity ratio at the systemic velocity}
\item[$^\mathrm{b}$] \footnotesize{Clump mass estimated from the C$^{18}$O(2-1) emission using the area under the whole emission line, and the spatial area delimited by the half power contour. $T_{\mathrm{ex}}$ assumed to be between 20 and 80~K}
\item[$^{\mathrm{c}}$] \footnotesize{Virial mass, computed using Eq. (6) of Fontani et al. (\cite{fontani02})}
\item[$^\mathrm{d}$] \footnotesize{Clump mass derived from the (sub)mm dust emission measured by the cited authors}
\item[$^\mathrm{e}$] \footnotesize{Rotational velocity; projected values (i.e. lower limits)}
\item[$^{\mathrm{f}}$] \footnotesize{Dynamical mass of the clump, estimated using Eq. (\ref{mdyn})}
\item[] \footnotesize{References: $^1$ Hill et al. (\cite{hill05}), $^2$ Fa\'undez et al. (\cite{faundez04}), $^3$ Mooney et al. (\cite{mooney95}), $^4$ Hatchell et al. (\cite{hatchell00}), $^5$ Mueller et al. (\cite{mueller02})}
\end{list}
\end{table*}

From the emission integrated under the \c18o line corrected for $\tau_{\mathrm{C^{18}O}}$, the clump mass, $M_{\mathrm{clump}}$, has been determined assuming an excitation temperature, $T_{\mathrm{ex}}$, between 20 and 80~K, and a C$^{18}$O abundance dependent on the Galactocentric distance (Wilson \& Rood \cite{wilson94}):

\begin{equation}
\frac{[\mathrm{C}^{16}\mathrm{O}]}{[\mathrm{C}^{18}\mathrm{O}]} = 58.8 d_\mathrm{gal} + 37.1
\label{ab18}
\end{equation}

We have adopted [CO]/[H$_2$]~$=10^{-4}$. Fractionation and selective photo-dissociation have been neglected. Table \ref{gauss} presents the results from the gaussian fit to the \c18o spectrum at the peak position of each integrated map (grey scale in Fig. \ref{outflows}). The line wings were ignored in the fit. The virial mass, $M_{\mathrm{vir}}$, of each clump, has been determined from the FWHM thus estimated, and from the angular diameter of the clump, using Eq. (6) of Fontani et al. (\cite{fontani02}). The values obtained for $M_{\mathrm{clump}}$ and $M_{\mathrm{vir}}$ are listed in Table \ref{mass}. It is evident, from a comparison between these two variables, that the virial masses are roughly of the order of the clump masses, indicating the molecular clumps are in virial equilibrium.

For comparison with the C$^{18}$O and virial masses, we have taken from the literature observations mapping the mm or submm continuum flux and we have converted these (see col.7 of Table \ref{mass}) into estimates of the clump mass inferred from the dust emission, $M_\mathrm{dust}$. For this purpose, we used the equation:

\begin{equation}
M_{\mathrm{dust}}=\frac{S_{\nu} d^2}{\kappa_{\nu} B_{\nu}(T_{\mathrm{d}}) R_{\mathrm{d}}}
\label{mdust}
\end{equation}
where $S_{\nu}$ is the measured flux, $d$ the distance to the source (see Table \ref{sample}), $\kappa_{\nu}$ the frequency-dependent dust mass opacity coefficient, $B_{\nu}(T_{\mathrm{d}})$ is the Planck function for a blackbody of dust temperature $T_{\mathrm{d}}$, and $R_{\mathrm{d}}$ is the dust-to-gas ratio, which we assume equal to 0.01. We have assumed $\kappa_{\nu}=\kappa_0 (\nu/\nu_0)^{\beta}$, with $\kappa_0=1$~cm$^2$~g$^{-1}$ at $\nu_0=250$~GHz (Ossenkopf \& Henning \cite{ossen94}), and $\beta = 2$, as in Beuther et al. (\cite{beut02b}). $T_{\mathrm{d}}$ varies from clump to clump according to the authors who measured the dust emission (see Table \ref{mass}). We note that no (sub)mm observations towards G75.78 are available in the literature, and that the dust mass obtained for G48.61 has been derived from 350~$\mu$m data. The issue of clump masses will be further discussed in Sect.~5.

\begin{table}[!hbt]
\caption{Results from gaussian fit to C$^{18}$O(2-1) peak spectra}
\label{gauss}
\centering
\begin{tabular}{lcccc}
\hline
Source & $T_{\mathrm{MB}}$ & FWHM & $V_{\mathrm{LSR}}$ & $\int T_{\mathrm{MB}}$~d$V$\\
 & (K) & (km~s$^{-1}$) & (km~s$^{-1}$) & (K~km~s$^{-1}$)\\
 \hline
G10.47A & 13.1 (3.1) & 6.6 (0.3) & 66.4 (0.1) & 92.1 (3.8)\\
G10.47B & 7.4 (1.0) & 4.9 (0.4) & 71.1 (0.2) & 38.6 (2.5)\\
G10.62 & 31.3 (1.7) & 6.8 (0.1) & $-$3.2 (0.1) & 227 (2)\\
G16.59 & 13.6 (1.3) & 3.9 (0.1) & 59.0 (0.1) & 56.2 (1.2)\\
G19.61 & 14.4 (1.8) & 6.5 (0.2) & 41.5 (0.1) & 99.8 (2.5)\\
G23.44 & 14.0 (3.2) & 3.9 (0.3) & 100.8 (0.1) & 58.0 (3.5)\\
G28.87 & 15.6 (1.6) & 3.7 (0.1) & 103.3 (0.1) & 60.9 (1.7)\\
G29.96 & 21.0 (1.4) & 3.8 (0.1) & 97.6 (0.1) & 84.9 (1.4)\\
G35.20 & 10.1 (1.1) & 3.7 (0.1) & 34.9 (0.1) & 40.4 (1.2)\\
G43.89 & 8.0 (2.0) & 4.8 (0.4) & 53.9 (0.1) & 41.1 (2.4)\\
G48.61A & 12.2 (2.3) & 3.8 (0.2) & 18.6 (0.1) & 49.7 (2.4)\\
G48.61B & 3.4 (0.7) & 5.6 (0.6) & 15.8 (0.2) & 20.4 (1.8)\\
G75.78 & 10.7 (2.4) & 5.0 (0.3) & $-$0.2 (0.1) & 56.8 (3.0)\\
 \hline
\end{tabular}
\end{table}

\begin{figure*}[!bht]
\centering
\includegraphics[angle=-90,width=18cm]{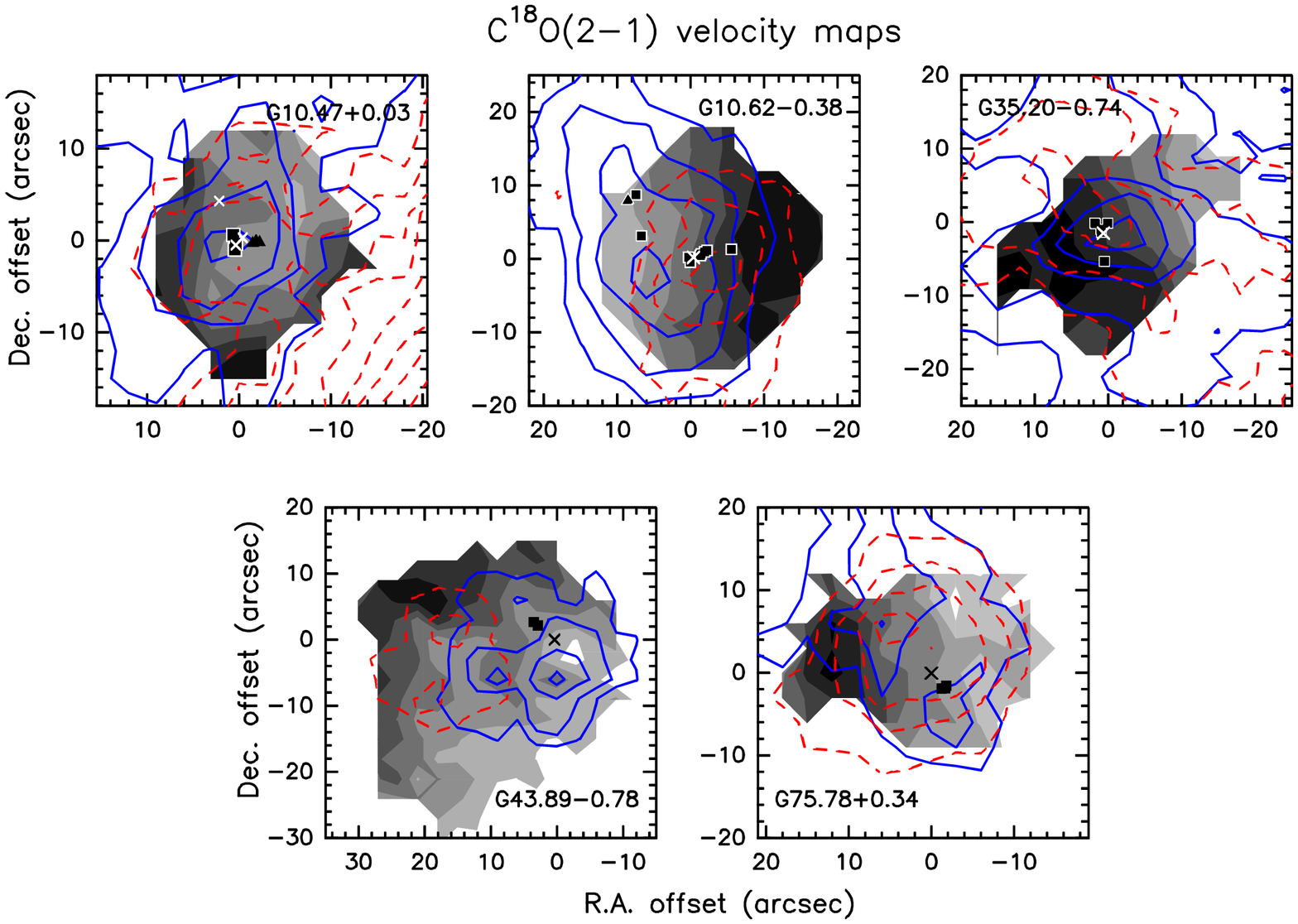}
\caption{Velocity gradients: the grey scale represents the velocity of the C$^{18}$O gas by steps of 0.2~km~s$^{-1}$. The bluest (white) and reddest (dark) levels correspond to 66.6 and 67.8~km~s$^{-1}$ (G10.47), $-$3.35 to $-$2.15~km~s$^{-1}$ (G10.62), 33.45 and 34.75~km~s$^{-1}$ (G35.20), 54.1 and 55.3~km~s$^{-1}$ (G43.89), $-$0.7 and 0.7 (G75.78)~km~s$^{-1}$. The blue (solid) and red (dashed) contours of the \13co outflow maps are overlaid for comparison. The symbols have the same meaning as in Fig. \ref{outflows}.}
\label{velo}
\end{figure*}

In some cases we detect velocity gradients in the \c18o emission (Fig. \ref{velo}). The interpretation of these is not straightforward, but the comparison between the direction of the velocity gradient and that of the main outflow axis can give some hints. If the velocity gradient is parallel to the outflow axis, it could be reflecting the molecular outflow already observed in the \13co emission. On the other hand, if this gradient is perpendicular to the outflow axis, as is the case of, e.g., G35.20, it could be interpreted as rotation, similar to what is found in G24.78+0.08 and G31.41+0.31 (Beltr\'an et al. 2004). 

Whichever the nature of the velocity gradients, it cannot be firmly confirmed with single-dish observations alone. Complementary interferometric observations are necessary to assess their nature. However, under the assumption that there are rotating structures, we have estimated a dynamical mass, $M_{\mathrm{dyn}}$, for the five cases in which we detect a C$^{18}$O velocity gradient, from the equation:

\begin{center}
\begin{equation}
M_{\mathrm{dyn}}=\frac{V_{\mathrm{rot}}^{2}R}{G}
\label{mdyn}
\end{equation}
\end{center}
where $V_{\mathrm{rot}}$ is the rotation velocity, $R$ the radius of the supposed disk, and $G$ the universal gravitational constant. This equation assumes  equilibrium between centrifugal and gravitational forces, and yields only a lower limit to the actual dynamical mass because $V_{\mathrm{rot}}$ is not corrected for inclination. Both $V_{\mathrm{rot}}$ and $M_{\mathrm{dyn}}$ are presented in Table \ref{mass}. Our values are small compared to the other clump mass estimates, and of the same order of magnitude as those found by Beltr\'an et al. (\cite{beltran04}). 

\section{Derived outflow parameters}

\subsection{Outflow velocity ranges}

The determination of outflow parameters from our $^{13}$CO spectra is non-trivial. The first complication arises when trying to select the outflow velocity ranges: there are no universal methods to separate the outflow from the ambient gas. If one includes too low-velocities there might be contamination from the ambient gas and the mass of the outflow may be overestimated. On the contrary, if one prefers to be safe from including ambient gas emission and chooses only the highest velocities, an important part of the outflow mass may be missed. 

The strategy we have used to select the blue and red velocity ranges of the outflows is the following. From the \13co\ and \c18o\ spectra in Fig. \ref{spec2}, we have defined the low-velocity limits where the \c18o line reaches a level of 2$\sigma$ or, in some cases, the beginning of weak non-gaussian wings. The high-velocity limits have been chosen where the \13co falls below 2$\sigma$. With the extra aid of the channel maps, these limits were either conserved or slightly modified depending on the blue- and red-shifted emission spatial distribution observed. The resulting blue and red velocity ranges, $\Delta V_{\mathrm{blue}}$ and $\Delta V_{\mathrm{red}}$, are listed in Table \ref{outcal}. These ranges have been used both to make the outflow maps in Fig. \ref{outflows} and to calculate the outflow parameters. For G10.47, no $\Delta V_{\mathrm{red}}$ is shown because the red-shifted outflow emission is contaminated by a second clump located $\sim$30\arcsec\ towards the south-west. Similarly, no $\Delta V_{\mathrm{blue}}$ is given for G28.87 because the blue emission does not show a well-defined structure in the map.

\begin{table*}[!tbh]
\caption{Velocity range of the \13co wings used for the calculations and lobe sizes}
\label{outcal}
\centering
\begin{tabular}{ccccccc}
\hline
Source & $\Delta V_{\mathrm{blue}}$ & $\Delta V_{\mathrm{red}}$ & $V_{\mathrm{max_{\mathrm{b}}}}$ & $V_{\mathrm{max_{\mathrm{r}}}}$ & Size$_{\mathrm{blue}}$ & Size$_{\mathrm{red}}$\\
 & (km~s$^{-1}$) & (km~s$^{-1}$) & (km~s$^{-1}$) & (km~s$^{-1}$) & (pc) & (pc)\\
\hline
G10.47 & [56.3, 60.8] & --- & 10.1 & --- & 2.4 & ---\\
G10.62 & [$-$11.6, $-$8.6] & [3.9, 7.9] & 8.4 & 11.1 & 0.9 & 0.7\\
G16.59 & [55.0, 56.5] & [62.0, 64.0] & 4 & 5 & 0.5 & 0.5\\
G19.61 & [29.1, 35.1] & [48.6, 50.6] & 12.4 & 9.1 & 0.3 & 1.2\\
G28.87 & --- & [108.1, 113.6] & --- & 10.3 & --- & 0.8\\
G35.20 & [25.3, 30.3] & [37.8, 43.8] & 9.6 & 8.9 & 1.4 & 0.8\\
G43.89 & [41.7, 49.7] & [57.2, 61.2] & 12.2 & 7.3 & 0.3 & 0.4\\
G48.61 & [9.7, 11.2] & [21.7, 25.2] & 8.9 & 6.6 & 0.8 & 0.9\\
G75.78 & [$-$9.1, $-$4.1 & [4.4, 10.9] & 8.9 & 11.1 & 0.5 & 0.4\\
\hline
\end{tabular}
\end{table*}  

\subsection{Optical depth}

Thanks to the high signal-to-noise level of the observations in the \13co\ wings and even in most of the \c18o line wings, it has been possible to estimate the optical depth of the \13co outflow emission (which we denote as $\tau_{\mathrm{blue}}$ and $\tau_{\mathrm{red}}$ for the blue and red emission, respectively) as done for $\tau_{\mathrm{C^{18}O}}$ (see Sect. 3.2). One should realise that the S/N ratio in the \c18o wings is sufficiently high only for the lowest-velocity channel of each outflow velocity range, and therefore $\tau_{\mathrm{blue}}$ and $\tau_{\mathrm{red}}$ have been computed only for such channels. Consequently, the values we obtain should be taken as upper limits to the mean optical depth throughout the \13co wings, and the same applies to the derived outflow parameters (mass, momentum, etc). In the case of G28.87 and G43.89, the signal in the \c18o\ line wings was not high enough to provide an optical depth estimate. In Table \ref{tot1} we do not indicate $\tau_{\mathrm{blue}}$ and $\tau_{\mathrm{red}}$, but the mean value, $\tau_{\mathrm{m}}=(\tau_{\mathrm{blue}}+\tau_{\mathrm{red}})/2$. We note, however, that even though we will refer to $\tau_{\mathrm{m}}$ in the text from now on, it is actually $\tau_{\mathrm{blue}}$ and $\tau_{\mathrm{red}}$ that have been used in the calculations.

\subsection{Outflow energetics and kinematics}

For each outflow, the mass, momentum and energy ($M$, $p$ and $E$, respectively) have been derived from the \13co\ emission under each line wing, $\Delta V_{\mathrm{blue}}$ or $\Delta V_{\mathrm{red}}$, as follows:

\begin{equation}
M = \sum_{ch = 1}^{n} m_{ch}
\end{equation}

\begin{equation}
p = \sum_{ch = 1}^{n} m_{ch} |V_{ch}-V_{\mathrm{sys}}|
\end{equation}

\begin{equation}
E = \frac{1}{2} \sum_{ch = 1}^{n} m_{ch} (V_{ch}-V_{\mathrm{sys}})^{2}
\end{equation}
where the summation is for all the velocity channels within $\Delta V_{\mathrm{blue}}$ or $\Delta V_{\mathrm{red}}$, $V_{ch}$ is the velocity of the corresponding channel, $V_{\mathrm{sys}}$ the systemic velocity of the clump (from Table \ref{gauss}), and $m_{ch}$ is the mass measured in each channel within the area defined by the 5$\sigma$ contour level of the corresponding outflow lobe (i.e. the most external outflow contours in Fig. \ref{outflows}). The sum of the contributions from each pixel and velocity channel builds up the total outflow mass, momentum and energy. The abundance ratio [$^{13}$CO]/[H$_2$], as in the case of C$^{18}$O, depends on the Galactocentric distance of each source according to the following relation (Wilson \& Rood \cite{wilson94}):

\begin{equation}
\frac{[^{12}\mathrm{CO}]}{[^{13}\mathrm{CO}]} = 7.5 d_\mathrm{gal} + 7.6
\label{ab13}
\end{equation}
The excitation temperature assumed is $T_{\mathrm{ex}}=$~10~K (the numbers decrease by $\sim$~10\% if $T_{\mathrm{ex}}=$~20~K is used instead of 10~K, and coincide for $T_{\mathrm{ex}}=$~30~K). The total outflow mass, momentum and energy obtained are presented in Table \ref{tot1}, both for the case $\tau \ll$~1 and corrected for $\tau_{\mathrm{m}}$. On average, $\tau_{\mathrm{m}} \simeq 2$, which is not negligible and causes $M$, $p$ and $E$ to increase by a factor $\sim$2.5.

\begin{table}[!htb]
\caption{Total outflow energetics.}
\label{tot1}
\vspace{1mm}
\centering
\begin{tabular}{lccccccc}
\hline
Source & $\tau_{\mathrm{m}}$$^\mathrm{\dag}$ & \multicolumn{2}{c}{$M$} & \multicolumn{2}{c}{$p$} & \multicolumn{2}{c}{$E$}\\
 & &  \multicolumn{2}{c}{(M$_{\sun}$)} & \multicolumn{2}{c}{(M$_{\sun}$~km~s$^{-1}$)} & \multicolumn{2}{c}{($10^{46}$~erg)}\\
  & & $\tau \ll 1$ & $\tau_{\mathrm{m}}$ & $\tau \ll 1$ & $\tau_{\mathrm{m}}$ & $\tau \ll 1$ & $\tau_{\mathrm{m}}$\\
 \hline
 G10.47$^\mathrm{\ddag}$ & 3.6 & 150 & 570 & 1110 & 4130 & 8.1 & 30\\
 G10.62 & 1.9 & 90 & 200 & 670 & 1510 & 5.0 & 12\\
 G16.59 & 2.1 & 24 & 59 & 85 & 210 & 0.31 & 0.75\\
 G19.61 & 3.6 & 200 & 740 & 1740 & 6320 & 15 & 55\\
 G28.87$^\mathrm{\ddag}$ & --- & 70 & --- & 470 & --- & 3.3 & ---\\
 G35.20 & 1.6 & 110 & 230 & 610 & 1220 & 3.5 & 6.9\\
 G43.89 & --- & 62 & --- & 380 & --- & 2.5 & ---\\
 G48.61 & 0.9 & 590 & 900 & 2550 & 3870 & 11.8 & 18\\
 G75.78 & 0.9 & 140 & 220 & 920 & 1370 & 6.0 & 9.0\\
 \hline
\end{tabular}
\begin{list}{}{}
\item[$^\mathrm{\dag}$] \footnotesize{Mean $\tau$ of the \13co\ line over the blue and red wings}
\item[$^\mathrm{\ddag}$] \footnotesize{Only the blue lobe parameters, for G10.47, and the red lobe parameters, for G28.87, were determined, due to contamination from nearby unrelated clumps}
\end{list}
\end{table}

A kinematic timescale, $t_{\mathrm{kin}}$, for each flow lobe has been determined from the equation $t_{\mathrm{kin}}=r/V_{\mathrm{out}}$, where $r$ is the projected size of the outflow lobe (see Table \ref{outcal}), and $V_{\mathrm{out}}$ is the difference in absolute value between the flow velocity and the systemic velocity ($V_{\mathrm{max}}$ in Table \ref{outcal}). The numbers thus obtained for the blue and red lobes have been averaged to produce the kinematic timescales given in Table \ref{tot2}.

From $t_{\mathrm{kin}}$, the outflow parameters $\dot{M}$, $\dot{p}$ and $L_{\mathrm{mec}}$, have been derived for the blue and red lobes separately, and then added together to obtain the total values presented in Table \ref{tot2}.

\begin{table*}[!htb]
\caption{Total outflow timescales and kinematics.}
\label{tot2}
\vspace{1mm}
\centering
\begin{tabular}{lcccccccc}
 \hline
Source & $\tau_{\mathrm{m}}$$^\mathrm{\dag}$ & $t_{\mathrm{kin}}$ & \multicolumn{2}{c}{$\dot{M}$} & \multicolumn{2}{c}{$\dot{p}$} & \multicolumn{2}{c}{$L_{\mathrm{mec}}$}\\
 & & ($10^4$~yr) & \multicolumn{2}{c}{($10^{-3}$~M$_{\sun}$~yr$^{-1}$)} & \multicolumn{2}{c}{($10^{-3}$~M$_{\sun}$~km~s$^{-1}$~yr$^{-1}$)} & \multicolumn{2}{c}{(L$_{\sun}$)}\\
  & & & $\tau \ll 1$ & $\tau_{\mathrm{m}}$ & $\tau \ll 1$ & $\tau_{\mathrm{m}}$ & $\tau \ll 1$ & $\tau_{\mathrm{m}}$\\
 \hline
 G10.47$^\mathrm{\ddag}$ & 3.6 & 23 & 0.66 & 2.4 & 4.7 & 18 & 2.9 & 11\\
 G10.62 & 1.9 & 8.4 & 1.1 & 2.5 & 8.4 & 20 & 5.5 & 13\\
 G16.59 & 2.1 & 11 & 0.23 & 0.56 & 0.82 & 2.0 & 0.26 & 0.62\\
 G19.61 & 3.6 & 8.0 & 4.4 & 17 & 38 & 150 & 28 & 110\\
 G28.87$^\mathrm{\ddag}$ & --- & 8.1 & 0.85 & --- & 5.8 & --- & 3.4 & ---\\
 G35.20 & 1.6 & 11 & 1.0 & 2.0 & 5.2 & 10 & 2.5 & 5.0\\
 G43.89 & --- & 4.0 & 1.9 & --- & 12 & --- & 7.0 & ---\\
 G48.61 & 0.9 & 11 & 6.1 & 9.3 & 26 & 39 & 10 & 15\\
 G75.78 & 0.9 & 4.5 & 3.5 & 5.2 & 22.3 & 33 & 13 & 18\\
 \hline
\end{tabular}
\begin{list}{}{}
\item[$^\mathrm{\dag}$] \footnotesize{Mean $\tau$ of the \13co\ line over the blue and red wings}
\item[$^\mathrm{\ddag}$] \footnotesize{Only the blue lobe parameters, for G10.47, and the red lobe parameters, for G28.87, were determined, due to contamination from nearby secondary clumps}
\end{list}
\end{table*}

\subsection{Outflow inclination}

We have no means to determine from our data the inclination, $i$, of each outflow. Here we assume $i$ as the angle between the outflow axis and the line of sight. Therefore the parameters we have derived so far (Tables \ref{tot1} and \ref{tot2}) are subject to an uncertainty. As an indication, in Table \ref{geom} we show the range of values for the outflow parameters after correcting for $i$ of between 30\degr and 60\degr. Quantities that depend strongly on the flow velocity, i.e. the kinematic energy and the mechanical luminosity, are most affected by the inclination, especially at high values of $i$. The numbers variate from a factor $\sim$0.6 for $\dot{M}$ at 30\degr to a factor $\sim$7 for L$_\mathrm{mec}$ at 60\degr.

\begin{table*}[!htb]
\caption{Total outflow parameters after correction for an inclination angle in the range 30$^{\circ} - 60^{\circ}$ ($\tau_{\mathrm{m}}$ taken into account)}
\label{geom}
\centering
\begin{tabular}{lccccc}
\hline
Source & $p$ & $E$ & $\dot{M}$ & $\dot{p}$ & $L_{\mathrm{mec}}$\\
 &  (M$_{\odot}$~km~s$^{-1}$) & 10$^{46}$~erg & ($10^{-3}$M$_{\odot}$~yr$^{-1}$) & ($10^{-3}$~M$_{\odot}$~km~s$^{-1}$~yr$^{-1}$) & (L$_{\odot}$)\\
 \hline
G10.47 & 4770$-$8260 & 40$-$120 & 1.4$-$4.2 & 12$-$61 & 8.3$-$75\\
G10.62 & 1740$-$3010 & 15$-$46 & 1.4$-$4.3 & 13$-$68 & 10$-$92\\
G16.59 & 240$-$410 & 1.0$-$3.0 & 0.32$-$0.97 & 1.3$-$6.8 & 0.48$-$4.3\\
G19.61 & 7300$-$12600 & 73$-$220 & 9.8$-$29 & 97$-$500 & 82$-$740\\
G28.87$^\mathrm{\dag}$ & 540$-$940 & 4.3$-$13 & 0.49$-$1.5 & 3.9$-$20 & 2.6$-$24\\
G35.20 & 1410$-$2400 & 9.2$-$28 & 1.1$-$3.4 & 7.0$-$36 & 3.8$-$35\\
G43.89$^\mathrm{\dag}$ & 440$-$770 & 3.3$-$10 & 1.1$-$3.3 & 8.0$-$42 & 5.3$-$48\\
G48.61 & 4470$-$7740 & 24$-$71 & 5.3$-$16 & 26$-$140 & 12$-$100\\
G75.78 & 1590$-$2750 & 12$-$36 & 3.0$-$9.0 & 22$-$110 & 14$-$130\\
 \hline
\end{tabular}
\begin{list}{}{}
\item[$^\mathrm{\dag}$] \footnotesize{G28.87 and G43.89 have no $\tau_\mathrm{m}$ estimate}
\end{list}
\end{table*}

\section{Discussion}

\subsection{\element[][13]{CO} versus \element[][12]{CO}}

The \element[][12]{CO} molecule is about 77 times more abundant than \element[][13]{CO} in the interstellar medium (Wilson \& Rood \cite{wilson94}), and consequently optically thicker. In order to understand the effects of using one tracer or another on the derivation of outflow parameters, we have made the following analysis: we have compared our outflow parameters, obtained from our \13co\ observations, to those derived by other authors from \co\, for the sources G16.59$-$0.06 and G75.78$+$0.34. The former belongs to the sample surveyed by Beuther et al. (\cite{beut02a}) and has been studied in detail by the same authors more recently (Beuther et al. \cite{beut06}), both with IRAM~30-m and with the Sub Millimeter Array (SMA); the latter has been mapped at high angular resolution by Shepherd et al. (\cite{shepherd97}; see Appendix). Interferometric observations were also performed towards G16.59 in the \element[][12]{CO}(1$-$0) transition by Furuya et al. (\cite{furuya08}), revealing just one of the two outflows resolved by Beuther et al. (\cite{beut06}) with a combination of SMA and IRAM~30-m images, and so we do not use their results in this comparison. To avoid effects due to the uncertainties in the kinematic timescales of the outflows we use only $M$, $p$ and $E$ in the comparison, which are time-independent.

We note that Beuther et al. (\cite{beut06}) chose the same low-velocity limits as us in $\Delta V_{\mathrm{blue}}$ and $\Delta V_{\mathrm{red}}$, but used a $^{13}$CO abundance of 1.1$\times 10^{-6}$ relative to H$_{2}$, which in the case of G16.59 is smaller than the one we have used by a factor 2.3. Therefore, just for the purpose of this section, we have modified our G16.59 outflow parameters accordingly. On the other hand, Shepherd et al. (\cite{shepherd97}) chose low-velocity limits higher than ours, missing the lowest-velocity channels that we have included in our calculations. We have therefore re-computed our outflow parameters using the same limits as Shepherd et al. (\cite{shepherd97}) to allow for a coherent comparison. The numbers are presented in Table \ref{comp}, together with the tracers used by the cited authors.

\begin{table*}[!tbh]
\caption{Comparison of outflow mass, momentum and kinetic energy between this work and other studies}
\label{comp}
\centering
\begin{tabular}{ccccccccccc}
\hline
Source & Tracer & $M_{\mathrm{other}}$ & \multicolumn{2}{c}{$M_{\mathrm{our}}$} & $p_{\mathrm{other}}$ & \multicolumn{2}{c}{$p_{\mathrm{our}}$} & $E_{\mathrm{other}}$ & \multicolumn{2}{c}{$E_{\mathrm{our}}$}\\
 & & (M$_{\sun}$) & \multicolumn{2}{c}{(M$_{\sun}$)} & (M$_{\sun}$~km~s$^{-1}$) & \multicolumn{2}{c}{(M$_{\sun}$~km~s$^{-1}$)} & (10$^{46}$~erg) & \multicolumn{2}{c}{(10$^{46}$~erg)}\\
 & & & $\tau \ll$~1 & $\tau_{\mathrm{m}}$ & & $\tau \ll$~1 & $\tau_{\mathrm{m}}$ & & $\tau \ll$~1 & $\tau_{\mathrm{m}}$\\
\hline
G16.59$^\mathrm{\dag}$ & \element[][12]{CO}(2$-$1) & 33 & 55 & 130 & 268 & 190 & 480 & 2.5 & 0.71 & 1.7\\
G75.78$^\mathrm{\ddag}$ & \element[][12]{CO}(1$-$0) & 54 & 66 & 100 & 667 & 520 & 800 & 8.3 & 4.1 & 6.3\\
\hline
\end{tabular}
\begin{list}{}{}
\item[$^\mathrm{\dag}$] \footnotesize{$M_{\mathrm{other}}$, $p_{\mathrm{other}}$ and $E_{\mathrm{other}}$ from single-dish observations by Beuther et al. (\cite{beut02a}, \cite{beut06}), corrected for revised distance}
\item[$^\mathrm{\ddag}$] \footnotesize{$M_{\mathrm{other}}$, $p_{\mathrm{other}}$ and $E_{\mathrm{other}}$ from interferometric observations by Shepherd et al. (\cite{shepherd97}), corrected for revised distance; our outflow parameters re-calculated using the same low-velocity limits as Shepherd et al. (\cite{shepherd97})}
\end{list}
\end{table*}

From this table it is clear that, when the $\Delta V$ chosen are the same, all our parameters are comparable to those derived by Beuther et al. (\cite{beut06}) and Shepherd et al. (\cite{shepherd97}). Our mass estimates are somewhat higher, even when $\tau \ll$~1 is assumed. This suggests that the optical depth of the \element[][12]{CO} line wings may have been underestimated by the cited authors.  It is also true that our $\tau_{\mathrm{m}}$ have likely been overestimated for the reasons described in Sect. 4.2. That is why we also present our results for the optically thin case in Tables \ref{tot1} and \ref{tot2}, constraining our outflow parameter estimates to fall between the two limits.

On the other hand, our energy estimates are smaller on average than those derived by Beuther et al. (\cite{beut06}) and Shepherd et al. (\cite{shepherd97}). This is due to the more extended wings of the optically thicker \element[][12]{CO} lines. Unlike the mass, the kinetic energy estimate is most sensitive to the higher velocity channels (see discussion by Masson \& Chernin \cite{masson94}), and so it is to be expected that a line with broader wings will give rise to higher kinetic energies. Moreover, the effect of the optical depth is much smaller in this case because it is negligible in the high-velocity channels.

A compromise between the mass and the energy is the momentum of the flows, which does not depend so much on the extremes of the velocity ranges chosen (again, see Masson \& Chernin \cite{masson94}). Indeed, we see from Table \ref{comp} that $p$ is the parameter that shows the smallest discrepancy between different studies.

We conclude from this comparative analysis that the use of the \element[][13]{CO} isotopologue will generally produce more reliable masses, similar momentum and underestimated kinetic energies relative to what is obtained using \element[][12]{CO}. This effect is evident in Fig. \ref{lum3} (see Sect. 5.2)

Note that the loss of the highest velocity channels when using \element[][13]{CO} instead of \element[][12]{CO} also leads to higher kinematic timescales. Indeed, the values of $t_\mathrm{kin}$ obtained in the present work are on average $\sim$1.5 times higher than those derived by Beuther et al. (\cite{beut02a}), a result which can be explained by the fact that our mean outflow $V_\mathrm{max}$ is a factor $\sim$1.5 lower.

\subsection{Outflow parameters as a function of luminosity}

In this section, we discuss our results and compare them to those obtained by Beuther et al. (\cite{beut02a}), who carried out a similar survey in less luminous sources.

\begin{figure}[!bht]
\centering
\includegraphics[width=8cm]{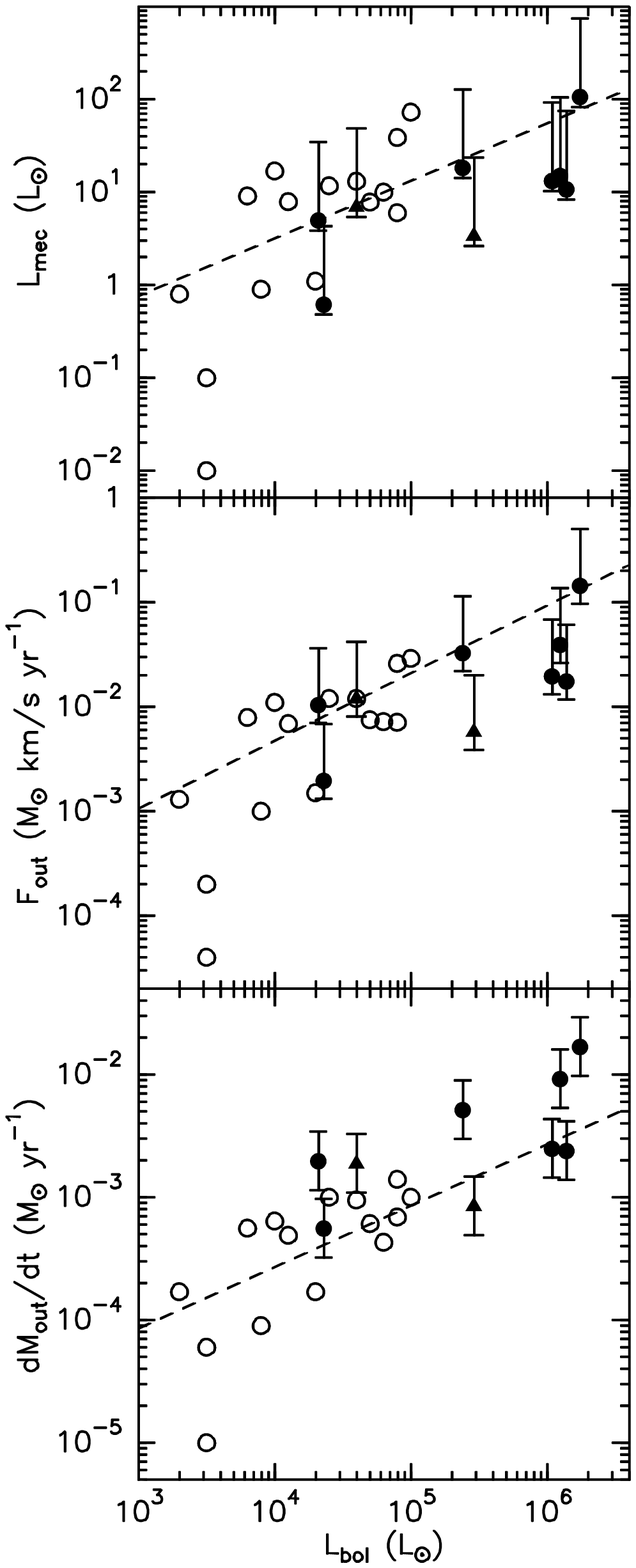}
\caption{\emph{Top}: Mechanical luminosity of the outflow versus bolometric luminosity. \emph{Middle}: Mechanical force of the outflow versus bolometric luminosity. \emph{Bottom}: Mass loss rate versus bolometric luminosity. Filled symbols correspond to the sources in our sample and open circles to data from Beuther et al. (\cite{beut02a}). The vertical bars indicate, respectively, the variation of $L_{\mathrm{mec}}$, $F_{\mathrm{out}}$ and $\dot{M}$ with inclination: the lower limit corresponds to an inclination of the outflow axis of 30\degr with respect to the line of sight, whereas the upper limit corresponds to an inclination of 60\degr. For our data, the optical depth is taken into account (circles; see Tables \ref{tot1} and \ref{tot2}). G28.87 and G43.89, for which no $\tau_{\mathrm{m}}$ estimate has been possible, are marked as filled triangles and represent lower limits to the outflow parameter plotted. The dashed lines represent the best fits to the data compiled by Wu et al. (\cite{wu04}, \cite{wu05}). Note that these are not fits to the data plotted in this figure.}
\label{lum3}
\end{figure}

\begin{figure}[!bht]
\centering
\includegraphics[width=8cm]{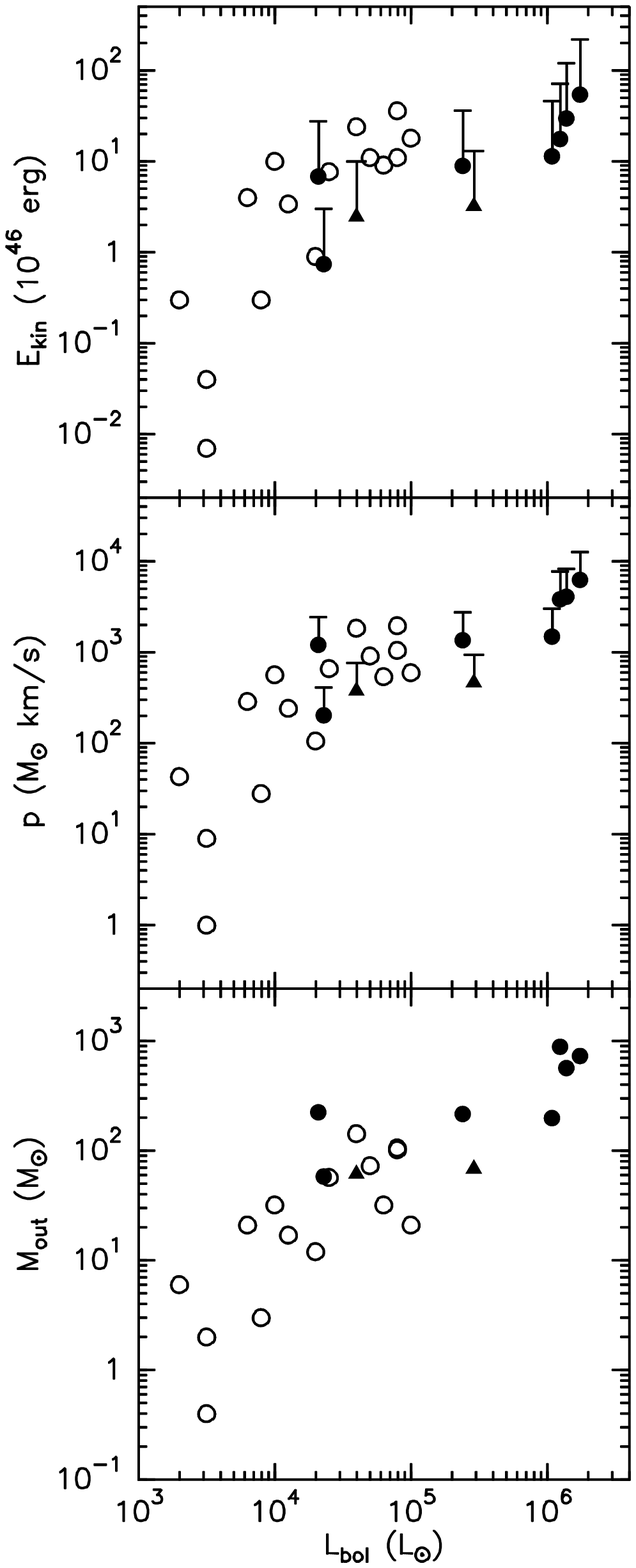}
\caption{\emph{Top}: Kinetic energy of the outflow versus bolometric luminosity. \emph{Middle}: Total outflow momentum of the outflow versus bolometric luminosity. \emph{Bottom}: Outflow mass versus bolometric luminosity. Filled symbols correspond to the sources in our sample and open circles to data from Beuther et al. (\cite{beut02a}). The vertical bars indicate, respectively, the variation of $E_{\mathrm{kin}}$ and $p$ with inclination: it corresponds to an inclination of 60\degr with respect to the line of sight. For our data, the optical depth is taken into account (circles; see Tables \ref{tot1} and \ref{tot2}). G28.87 and G43.89, for which no $\tau_{\mathrm{m}}$ estimate has been possible, are marked as filled triangles and represent lower limits to the outflow parameter plotted.}
\label{lum3_not}
\end{figure}

In Fig. \ref{lum3}, the time-dependent outflow parameters, i.e. mass loss rate, $\dot{M}$, mechanical force, $\dot{p}$ or $F_{\mathrm{out}}$, and mechanical luminosity, $L_{\mathrm{mec}}$, are plotted against bolometric luminosity, $L_{\mathrm{bol}}$ (Table \ref{sample}). The quantities obtained by Beuther et al. (\cite{beut02a}), for their sources with no distance ambiguity, are represented as open circles in Fig. \ref{lum3}, and our values as filled symbols: circles for the sources with a $\tau_{\mathrm{m}}$ estimate, and triangles for G28.87 and G43.89, for which no optical depth has been determined in the \13co\ line wings. The latter represent lower limits to the outflow parameter values. The same symbology applies to Figs. \ref{lum3_not}, \ref{lyman} and \ref{rest3}. The error bars correspond to a correction for an inclination varying from 30\degr\ to 60\degr\ (see Table \ref{geom}). It is immediately apparent from these plots that our sources have higher bolometric luminosities than those studied by Beuther et al. by a factor $\sim$10, and thus complement their study.

The results by Beuther et al. (\cite{beut02a}) show continuity in the correlation between mechanical force, $F_{\mathrm{out}}$, and bolometric luminosity, $L_{\mathrm{bol}}$, from low-mass to high-mass outflows, with higher mechanical forces corresponding to higher luminosities. In this work we find that this trend extends up to luminosities higher than $10^{5}$~L$_{\sun}$, indicating that the outflow formation mechanism is present up to luminosities of the order of 10$^6$~L$_{\mathrm{\sun}}$, and that more luminous sources drive more powerful outflows. Similar correlations are found for the mass loss rate, $dM_{\mathrm{out}}/dt$, and for the mechanical luminosity, $L_{\mathrm{mec}}$, of the outflows. For the latter, there seems to be a saturation towards the highest luminosity sources, but this may just be a consequence of using \element[][13]CO instead of the \element[][12]CO molecule used by Beuther et al. (\cite{beut02a}; as explained in Sect. 5.1). The weakening of the correlation between $dM_{\mathrm{out}}/dt$ and $L_{\mathrm{bol}}$ towards the highest luminosities claimed by Beuther et al. (\cite{beut02a}) is not evident in our plot: higher luminosities yield higher mass loss rates, with no apparent saturation in the highest luminosity range.

The behaviour of the data plotted in Fig. \ref{lum3} is consistent with the results obtained by Wu et al. (\cite{wu04}, Figs. 6 and 7; \cite{wu05}, Fig. 13), who compiled all the outflow studies available before February 2003 and performed a statistical analysis of the outflow parameters for a sample of sources spanning a wide range of bolometric luminosities (10$^{-1}$-10$^{6}$~L$_{\sun}$). Despite the heterogeneity of the sample, their best fits for $\dot{M}_\mathrm{out}$, $F_\mathrm{out}$ and $L_\mathrm{mec}$ (vs $L_\mathrm{bol}$) fit nicely the points in Fig. \ref{lum3} (dashed lines). As pointed out by the cited authors, the fact that these correlations hold for such a broad range of luminosities suggests that the luminosity of the powering source determines the outflow energetics, and that the driving mechanisms are similar for all luminosities.

It is unclear from our observations which is the multiplicity of the outflows. The increasing $F_{\mathrm{out}}$ with luminosity could be interpreted both as a single outflow powered by a more luminous source, or by more than one outflow coming from less luminous sources, each of them contributing to the overall outflow energetics (see Appendix for some particular cases). Such outflow multiplicity is generally found to be around 3 or 4 when interferometric observations are available (e.g. Beuther et al. \cite{beut06}). If we assume that the highest luminosity clumps contain a few molecular outflows, the only effect on the plot would be to split each point into several points shifted by a small factor (3 or 4) towards lower $F_{\mathrm{out}}$ and $L_\mathrm{bol}$, so the overall appearance of the plots in Fig. \ref{lum3} would remain the same.

For completeness, we also present plots of the time-independent outflow parameters as a function of $L_\mathrm{bol}$ (Fig. \ref{lum3_not}), which show an analogous behaviour to that of Fig. \ref{lum3}. We feel that, based on the arguments described, the general behaviour of the outflow properties plotted in Figs. \ref{lum3} and \ref{lum3_not} is real and not affected by possible errors in the calculation of the outflow parameters. This belief is reinforced by the continuity displayed by our data with respect to that of Beuther et al. (\cite{beut02a}).

\begin{figure}[!hbt]
\centering
\includegraphics[angle=-90,width=8cm]{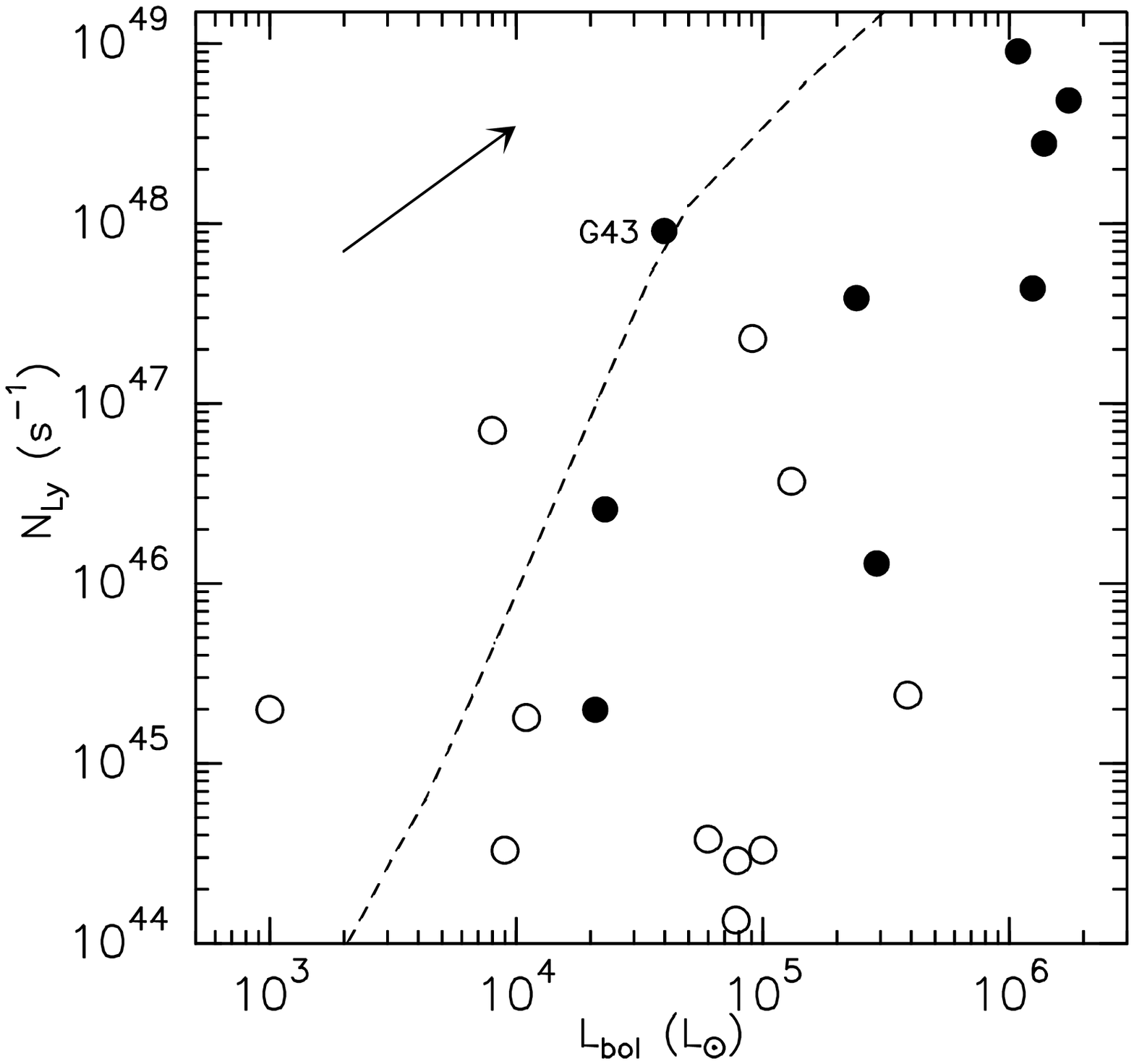}
\caption{Ionising photon rate, $N_\mathrm{Ly}$ against bolometric luminosity, $L_\mathrm{bol}$. Filled circles correspond to our data and open circles to data from Beuther et al. (\cite{beut02a}) and Sridharan et al. (\cite{srid02}). The dashed line represents the expected curve for an individual source on the ZAMS (Panagia \cite{pana73}). The arrow indicates the direction along which the points shift if the distance to the source is modified.}
\label{pana}
\end{figure}

In Fig. \ref{pana} we plot the Lyman photon rate of the associated UC \HII\ region(s) (see Table \ref{sample}), $N_{\mathrm{Ly}}$, against bolometric luminosity, $L_\mathrm{bol}$. We have computed $N_{\mathrm{Ly}}$ for the sources from Beuther et al. (\cite{beut02a}) which were detected in the VLA 3.6~cm survey by Sridharan et al. (\cite{srid02}), under the assumption that all the emission comes indeed from an \HII\ region and that it is optically thin. The dashed line depicts the expected curve for an individual source on the ZAMS (Panagia \cite{pana73}). The fact that most of the sources lie below the theoretical curve suggests that the observed regions contain a whole cluster of YSOs of different masses. All of them contribute to the total $L_\mathrm{bol}$, but only the most massive stellar objects, which power \HII\ regions, contribute to $N_{\mathrm{Ly}}$. G43.89 lies just above the theoretical ZAMS line, but as indicated by the arrow, an increase in the distance to this source would shift the point below the line. The same applies to the two other sources from Beuther et al. (\cite{beut02b}) sample which are also above the ZAMS line. It therefore becomes evident that an accurate distance determination is necessary in order to interpret the observations correctly.

In an attempt to assess our data as a function of a quantity which does not represent a whole cluster, such as $L_\mathrm{bol}$, but the individual massive stellar objects, in Fig. \ref{lyman} we plot $F_{\mathrm{out}}$ against the Lyman photon rate. There seems to be a correlation between $F_{\mathrm{out}}$ and $N_{\mathrm{Ly}}$ which holds within a range spanning from B2 to O6 stars, assuming the UC \HII\ regions are powered by a single stellar object. 

\begin{figure}[!hbt]
\centering
\includegraphics[angle=-90,width=8cm]{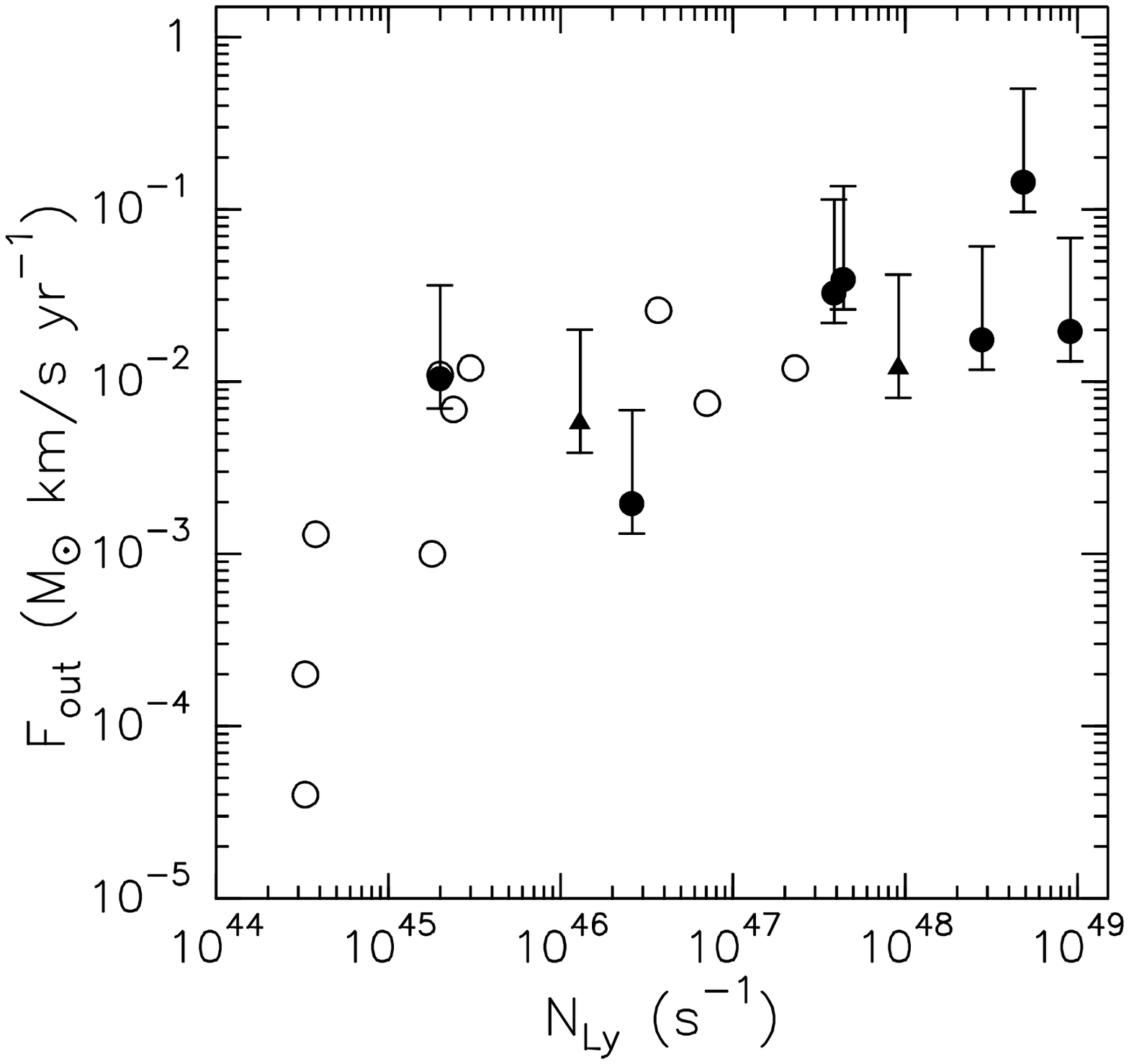}
\caption{Mechanical force of the outflow versus Lyman photon rate of the UCHII region(s) associated. Filled circles (corrected for $\tau_\mathrm{m}$) and triangles (with no $\tau_\mathrm{m}$ estimate) correspond to the sources in our sample and open circles to data from Beuther et al. (\cite{beut02a}) and Sridharan et al. (\cite{srid02}). The vertical bars indicate the variation of $F_{\mathrm{out}}$ with inclination: the lower limit corresponds to an inclination of the outflow axis of 30$^{\circ}$ with respect to the line of sight, whereas the upper limit corresponds to an inclination of 60$^{\circ}$.}
\label{lyman}
\end{figure}

Despite being less affected by source multiplicity, this plot displays a similar correlation as $F_{\mathrm{out}}$($L_{\mathrm{bol}}$) in Fig. \ref{lum3}, and indicates that more massive objects power more energetic outflows. Even if there are multiple UC \HII\ regions contributing to the total ionising photon rate, the good correlation between $F_{\mathrm{out}}$ and $N_{\mathrm{Ly}}$ shown by this plot would suggest that each outflow can be associated with an individual massive stellar object.

\subsection{Outflow parameters versus clump mass}

\begin{figure}[!tbh]
\centering
\includegraphics[width=8.5cm]{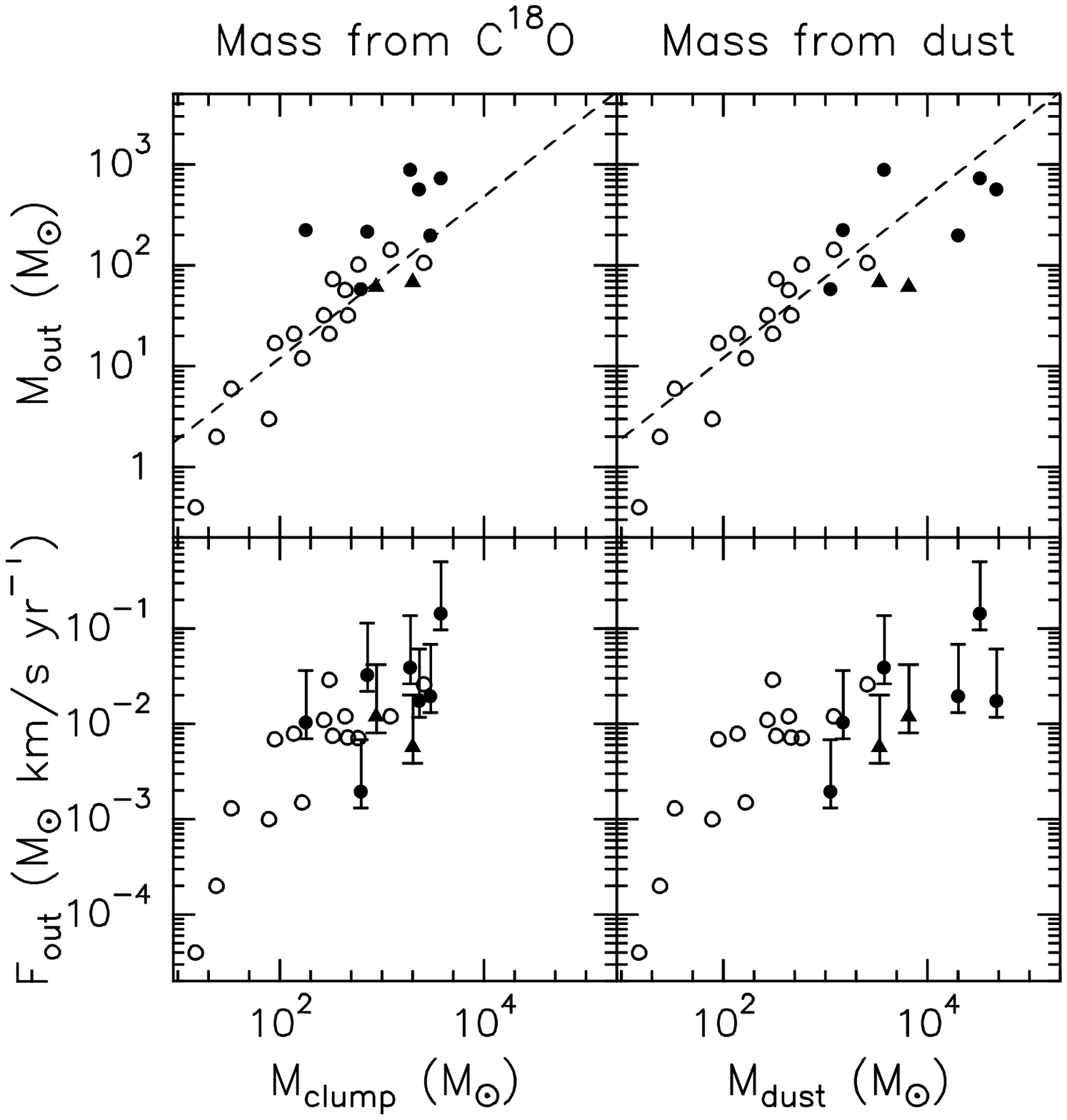}
\caption{\emph{Top panels}: Total molecular outflow mass versus clump mass from our \c18o data (left) and from continuum (sub)millimetric surveys by the authors cited in Table \ref{mass} (right). The dashed line is the best fit obtained by Beuther et al (\cite{beut02a}) for their data. \emph{Bottom panels}: Mechanical force of the outflow versus core mass as derived from the \c18o emission (left) and from continuum (sub)millimetric surveys by other authors (right). The vertical bars indicate the variation of $F_{\mathrm{out}}$ with inclination: the lower limit corresponds to an inclination of the outflow axis of 30\degr with respect to the line of sight, whereas the upper limit corresponds to an inclination of 60\degr. Filled circles represent the sources in our sample and open circles data from Beuther et al. (2002a), for which clump masses correspond to data from continuum emission at 1.2~mm in all four panels. For our data, the optical depth is taken into account (circles; see Tables \ref{tot1} and \ref{tot2}). G28.87 and G43.89, for which no $\tau_{\mathrm{m}}$ estimate has been possible, are marked as filled triangles and represent lower limits to the outflow parameter values.}
\label{rest3}
\end{figure}

Another way of inspecting our data is to plot some outflow parameters as a function of clump mass. Figure \ref{rest3} shows outflow mass, $M_{\mathrm{out}}$, and mechanical force, $F_{\mathrm{out}}$, against clump mass. On the left panels, our clump mass is the one obtained from the \c18o emission (see Table \ref{mass}), while masses of Beuther et al. (\cite{beut02a}) were determined from the 1.2~mm continuum emission of the clumps (Beuther et al. \cite{beut02b}). Therefore, for consistency in the comparison, we show on the right panels the same plots, with the only difference that our clump masses here correspond to $M_{\mathrm{dust}}$ in Table \ref{mass} (see Sect. 3.2). In order to homogenize the two samples, we have re-computed Beuther's clump masses from their measured 1.2~mm~fluxes (Beuther et al. \cite{beut02b}) using Eq. (\ref{mdust}), and we find them to be smaller by a factor 4.6 with respect to those reported in the original paper. This difference is due both to our lower assumed grain opacity with respect to the one used by Beuther et al., and to an error in the grain emissivity approximation by the same authors (see their Erratum, published in 2005). Once more, it is evident that our data complement the data from Beuther et al. (\cite{beut02a}), adding higher mass molecular clumps.

Two main ideas can be extracted from an overall look at the left and right panels in Figure \ref{rest3}. Firstly, our dust clump masses are larger than the corresponding C$^{18}$O masses by an average factor of $\sim$5. This difference might be explained by the fact that C$^{18}$O and the (sub)mm continuum are tracing different parts of the clump. Indeed, different sizes of the clump are measured in each tracer: the angular FWHM measured in the (sub)mm surveys is larger than in our \c18o maps by a factor of $\sim$~2.5. This may account for great part of the difference between the two mass estimates. In addition, Hofner et al. (\cite{hofner00}) concluded from their survey that the masses derived from (sub)mm dust emission tend to be systematically higher than masses derived from \c18o\ and C$^{17}$O(2$-$1) lines by a factor 2. They point out that several sources of uncertainty may contribute to this discrepancy, such as C$^{18}$O abundance, optical depth estimates, and the dust grain emissivity adopted.

The second feature to note in Fig. \ref{rest3} is the better continuity from Beuther's data towards our higher mass data when $M_{\mathrm{dust}}$ is used (right panels). This continuity is not so clear when we plot our data using the \c18o clump mass instead (left panels), especially for the $M_{\mathrm{out}}$ plot. This is not surprising, since we are comparing parameters derived from different tracers. 

We find that when $M_{\mathrm{dust}}$ is considered, one sees a tight correlation between this and the mass of the outflow, $M_{\mathrm{out}}$. The approximate relation $M_{\mathrm{out}}=0.3 M_{\mathrm{dust}}^{0.8}$, which is the best fit to the data of Beuther et al. (\cite{beut02a}), is also a fair fit to our more massive molecular clumps (see Fig. \ref{rest3}).

\begin{figure}[!htb]
\centering
\includegraphics[angle=-90,width=8cm]{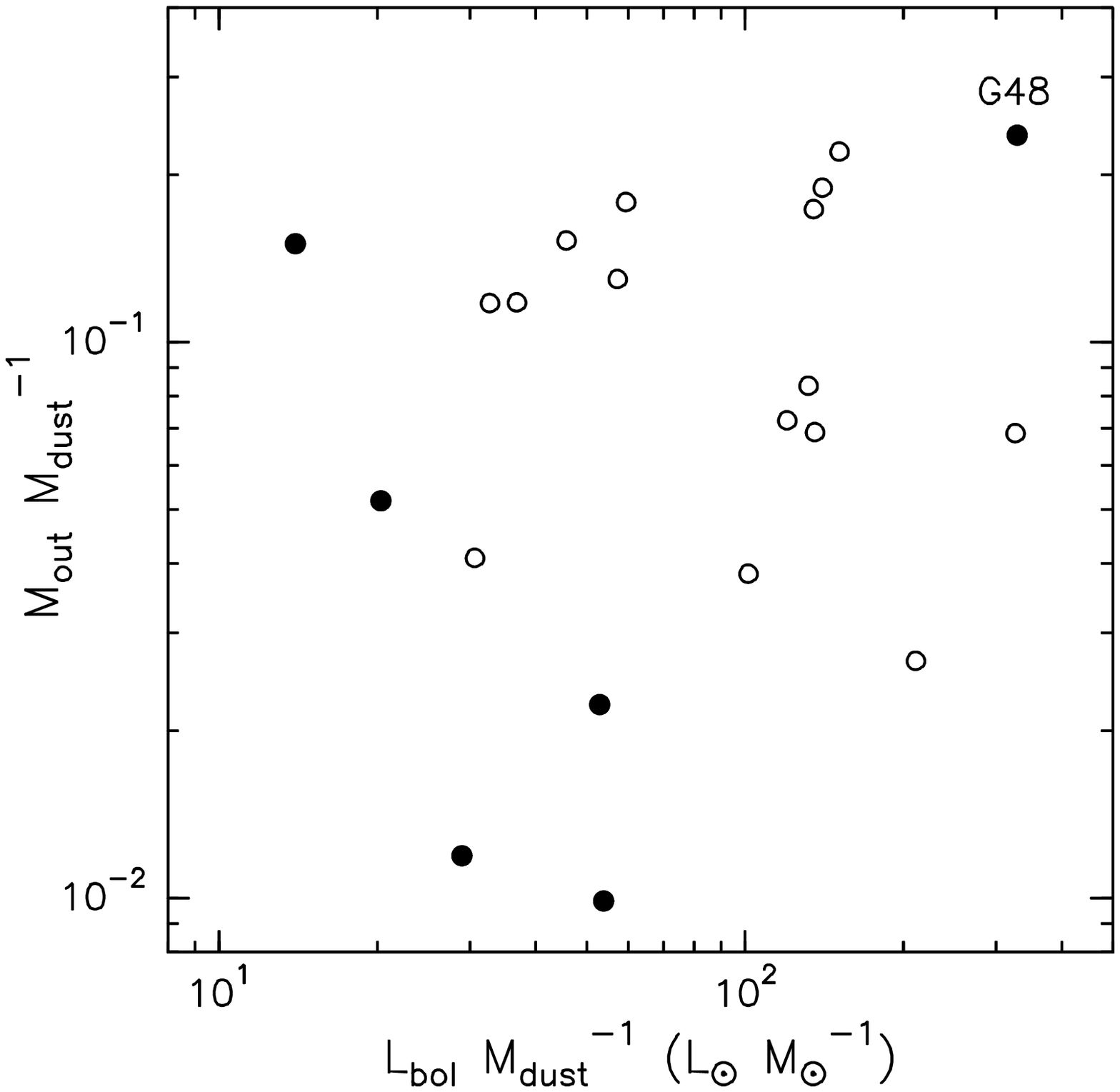}
\caption{$M_{\mathrm{out}}/M_{\mathrm{clump}}$ versus $L_{\mathrm{bol}}/M_{\mathrm{clump}}$. Filled circles correspond to the sources in our sample (only those with a $\tau_\mathrm{m}$ estimate are represented) and open circles to data from Beuther et al. (\cite{beut02a}).}
\label{imp}
\end{figure}

In Fig. \ref{imp}, we present a plot of $M_{\mathrm{out}}/M_{\mathrm{dust}}$ against $L_{\mathrm{bol}}/M_{\mathrm{dust}}$, which are distance-independent quantities. Ignoring G48.61, which is the only source for which the dust mass has been estimated from 350~$\mu$m data (see Sect. 3.2), two main interesting features characterise this plot. Firstly, our points tend to cluster to the left of the points representing Buether et al. This indicates that the most massive clumps tend to be ``under-luminous''. Secondly, and most importantly, the points span only a factor $10$ in $M_{\mathrm{out}}/M_{\mathrm{dust}}$, independently of their luminosity. This would suggest, consistently with Fig. \ref{rest3}, that the outflow mass is more strictly related to the mass of the clump than to the luminosity. In other words, the mass entrainment of the outflow is proportional to the mass of the whole clump.

\subsection{Accretion rates}

It is not straightforward to derive mass accretion rates from outflow mass loss rates. Nevertheless, by making a few assumptions it is possible to provide a rough estimate of the average accretion rate, $\dot{M}_\mathrm{accr}$, for our sample. Following the reasoning described in Beuther et al. (\cite{beut02a}), we consider the molecular outflows to be momentum driven. Therefore:

\begin{equation}
p_\mathrm{out} = p_\mathrm{jet} = \dot{M}_\mathrm{jet} v_\mathrm{jet} t_\mathrm{kin}
\end{equation}
Here we have assumed that the kinematic timescale of the jet is the same as that of the molecular outflow (listed in Table \ref{tot2}), which in our case is of order 10$^5$~yr. The average $p_\mathrm{out}$, (from Table \ref{tot1}) corrected for a mean outflow inclination of 45$\degr$, is $\sim$4000~M$_{\sun}$~km~s$^{-1}$. Adopting a typical jet velocity of about 700~km~s$^{-1}$ (e.g. Mart\'i et al. \cite{marti}), the mass loss rate of the jet, $\dot{M}_\mathrm{jet}$, is estimated to be on the order of 6$\times$10$^{-5}$~M$_{\sun}$~yr$^{-1}$ for sources with $L_\mathrm{bol}$ in the range 10$^{5}$-10$^{6}$~L$_{\sun}$. If we assume that $\dot{M}_\mathrm{jet}$ is about one third of the accretion rate (Tomisaka \cite{tami}), $\dot{M}_\mathrm{accr}$ amounts to $\sim$2$\times$10$^{-4}$~M$_{\sun}$~yr$^{-1}$, which is consistent with other values previously reported for high-mass SFRs (e.g. Beuther et al. \cite{beut02a}, Zhang et al. \cite{zhang05}). 

\subsection{Velocity gradients}

For one of our sources displaying a velocity gradient in the \c18o\ emission, and assuming such gradient is due to rotational motions, it is possible to find out whether the specific angular momentum, $l$, is conserved at different scales. This source is G10.62, for which other studies have revealed a rotating structure at a sub-parsec scale. G35.20 has also been reported to contain a rotating envelope by Little et al. (\cite{little85}). However, the size of such rotating structure is roughly the same as the one found in this work ($\sim$0.2~pc), and so we are not able to assess the conservation of $l$ at smaller scales.

Using $V_\mathrm{rot}$ in Table \ref{mass} and a radius of 0.6~pc, we obtain $l \simeq 0.2$~km~s$^{-1}$~pc for G10.62. On a smaller scale ($\sim$0.1~pc), we derive $l \simeq 0.3$~km~s$^{-1}$~pc from the C$^{18}$O interferometric observations carried out by Ho et al. (\cite{ho94}), which is of the same order as the value obtained from our data. One can compare this value of $l$ with the initial value of the parental clump. Assuming that the initial clump angular momentum is due to the Galactic differential rotation, we find an angular velocity $\omega \sim$10$^{-15}$~s$^{-1}$ (Clemens \cite{clem}). For a 1000~M$_{\sun}$ clump formed from a $\sim$10~pc radius sphere with a density of 1~cm$^{-3}$, typical of the interstellar medium, the initial $l$ is hence $\sim$3~\kms~pc, an order of magnitude higher than the derived (sub)parsec-scale value.

To sum up, although this analysis is based only on one source, it seems that the specific angular momentum is roughly conserved at scales smaller than about 1~pc, but it is not conserved on larger scales. A similar result was found by Ohashi et al. (\cite{oha99}, see their Fig.~2) for a sample of low-mass protostars. In their case, the radius inside which $l$ is conserved is smaller by about an order of magnitude, which indicates that infall takes place within a smaller radius than in the case of high-mass clumps.

\section{Summary}

We have mapped 11 high-mass SFRs in \13co and \c18o with the IRAM-30m telescope (Spain), with the aim to search for molecular outflows and characterise them and their associated molecular clumps. The sample is composed by high luminosity (up to $\sim$10$^6$~L$_{\sun}$) massive young clumps in their earliest evolutionary phases.

Our main conclusions can be summarised as follows:

\begin{enumerate}
\item The whole sample shows high-velocity wings in the \13co\ spectra, indicative of outflowing motions. We have obtained outflow maps in 9 of our 11 sources, which display well-defined blue and/or red lobes. This result indicates that outflows are as common in high-mass SFRs as in low-mass SFRs. Furthermore, it suggests that massive star formation may commonly proceed by accretion, in a similar way as low-mass star formation.
\item From the \c18o emission, the clump masses have been derived, taking into account the optical depth. The masses obtained are high and comparable to the virial masses, but have been found to be systematically smaller than the corresponding masses computed from the mm continuum emision by a factor $\sim$5. This is largely due to the different sensitivities in the two tracers.
\item The outflow parameters have been determined from the \13co emission in the high-velocity wings, by using standard methods. The values obtained are corrected for optical depth, which we have been able to estimate from the \13co to \c18o ratio in the line wings. Our results complement those by Beuther et al. (\cite{beut02a}), adding more luminous sources by an order of magnitude on average. A comparison between our results and those of Beuther et al. (\cite{beut02a}) reveals that there is continuity in the behaviour of outflow-related quantities as we move to the higher luminosity sources in our sample: higher outflow-related values correspond to higher luminosity sources. This trend also agrees quantitatively with the correlations derived by Wu et al. (\cite{wu04}, \cite{wu05}) for a large sample of outflow sources spanning a wide range of luminosities (0.1-10$^6$~L$_{\sun}$).
\item The mass of the outflow and the clump mass are tightly correlated, and the approximate relation $M_{\mathrm{out}}=0.3 M_{\mathrm{clump}}^{0.8}$, obtained by Beuther et al. (\cite{beut02a}) fits adequately the observations over three orders of magnitude.
\item We find a correlation between the outflow mechanical force, $F_{\mathrm{out}}$, and the rate of ionising photons, $N_{\mathrm{Ly}}$, of the UC \HII\ regions associated with the molecular clumps. High-angular resolution images are needed to resolve the outflows and associate them with individual sources within the clump.
\item Considering the molecular outflows in our sample to be momentum driven, we have estimated a mean accretion rate of $\sim 2 \times 10^{-4}$~M$_{\sun}$~yr$^{-1}$, which is on the order of what is typically found in high-mass SFRs (e.g. Beuther et al. \cite{beut02a}).
\item We find C$^{18}$O velocity gradients in 5 sources, in some cases they are compatible with rotating envelopes already detected with high-density tracers and at high-spatial resolution by other authors. The detection of velocity gradients on such large-scales with the C$^{18}$O molecule encourages further observations towards these five sources at higher-angular resolution and with high density tracers to understand their true nature.
\end{enumerate}

\begin{acknowledgements}
ALS acknowledges support from the FP6 Marie-Curie Research Training Network ``Constellation: the origin of stellar masses'' (MRTN-CT-2006-035890). NM is supported by Spanish MICINN through grants AYA2006-14876, by DGU of the Madrid community government under IV-PRICIT project S-0505/ESP-0237 (ASTROCAM), and by Molecular Universe FP6 MCRTN. It is a pleasure to thank the staff of IRAM Granada, who provided help during the observations. We are also grateful to the working group in Grenoble who have developed the free software package GILDAS, available at http://www.iram.fr/IRAMFR/GILDAS. We thank our referee, Henrik Beuther, for his valuable comments and suggestions which have helped to improve the scientific contents of this paper. 
\end{acknowledgements}

\begin{appendix}

\section{Comments on individual sources}

\subsection{G10.47+0.03}

With $L_{\mathrm{bol}} = 1.39 \times 10^{6}$~L$_{\sun}$, this is the most luminous source in our sample. This region contains 4 UC \HII\ regions, three of which are concentrated inside an area of 2\arcsec\ across inside a hot molecular core (Cesaroni et al.\cite{cesa98}), coinciding with \WAT\ masers (Hofner \& Churchwell, 1996). High-angular resolution C$^{18}$O maps obtained by Gibb et al. (\cite{gibb04}) reveal a velocity gradient in the SW-NE direction, which the authors interpret as a group of smaller individual cores with slightly different velocities. They conclude that there is weak evidence for either outflow or rotation in this region. On the other hand, high-resolution observations with PdBI by Olmi et al. (\cite{olmi96}) show an east to west CH$_3$CN velocity gradient, which they believe to be a rotating disk-like structure, and a south to north $^{13}$CO(1$-$0) velocity gradient which, added to the wide wings displayed by the line, could represent a molecular outflow.

\begin{figure}[!htb]
\centering
\includegraphics[width=8cm,angle=-90]{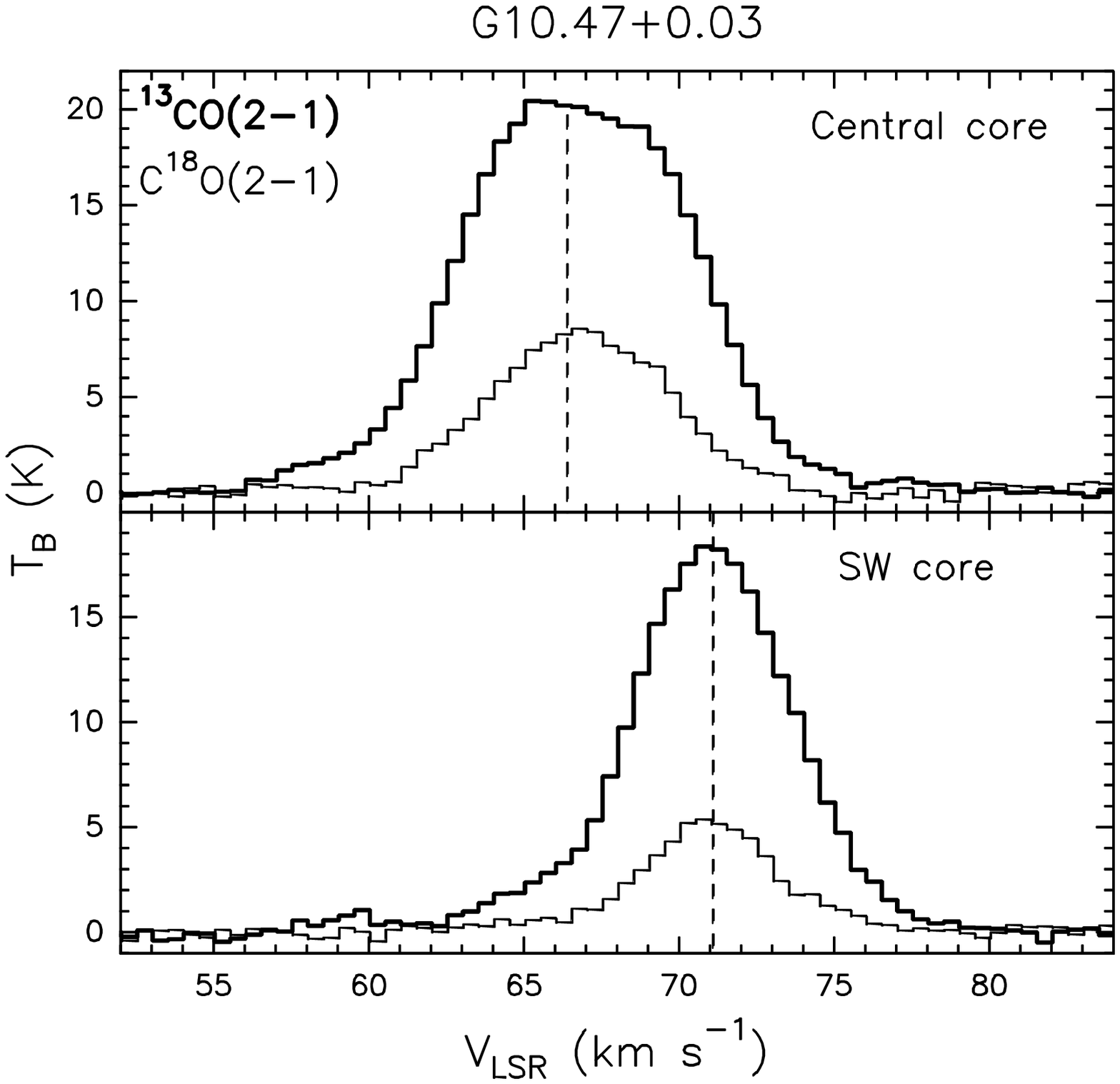}
\caption{Comparison between the \13co and \c18o spectra from the central core in the map of G10.47 (upper panel), and the corresponding spectra seen towards the secondary red-shifted core in the same map (lower panel; G10.46B core), located at an approximate offset of ($-30$\arcsec,$-25$\arcsec). The vertical dashed lines indicate the respective $V_{\mathrm{LSR}}$ listed in Table \ref{gauss}. Note the difference in their respective systemic velocities for both the \13co and \c18o lines.}
\label{g1047}
\end{figure}

On a larger scale, Klaassen \& Wilson (\cite{klaassen08}) report an HCO$^{+}$ outflow in the direction NE-SW, with a linear scale of $\sim$4~pc at 10.8~kpc. However, they stress that higher-resolution imaging is needed to confirm or reject this interpretation. In fact, from a close inspection of our maps we find that what they interpret as a molecular outflow is most likely a pair of cores with different systemic velocities (see Fig. \ref{g1047}). The contamination by the second core is actually the reason why we cannot obtain a clear red outflow lobe map from our single-dish observations.

As for the \c18o first moment, the velocity gradient seems the least reliable of the five shown in Figure \ref{velo}. Certainly the use of higher density tracers would be more adequate to assess the presence and nature of this velocity gradient.

\subsection{G10.62$-$0.38}

This source has been extensively studied in the past. It hosts an UC \HII\ region embedded in a hot molecular core (Keto et al. \cite{keto88}, Plume et al. \cite{plume92}). Its high Lyman photon rate (see Table \ref{sample}) implies an O6 spectral type for a ZAMS star, if the \HII\ region were powered by a single stellar object. However, VLA radio continuum maps performed by Sollins et al. (2005) reveal a clumpy structure of the ionised gas, suggesting the presence of multiple stars. 

Several authors have concluded from NH$_3$ observations that the HMC is rotating and collapsing inward onto and through the UC \HII\ region, and that rotation is seen at scales from 1~pc down to $\sim$0.05~pc (e.g. Sollins et al. \cite{sollins05} and references therein). The direction of the corresponding velocity gradient found by these authors goes from southeast (blue) to northwest (red) at small scales (2\arcsec). Our results show roughly an east to west velocity gradient in the \c18o line (Figure \ref{velo}), which despite having a slightly different orientation, could represent the large-scale continuation of the rotating core seen at high resolution by the above mentioned authors. Several \WAT\ maser spots are aligned along the direction of the \c18o velocity gradient.

Our \13co outflow map displays well defined lobes and suggests an outflow axis along a northeast-soutwest direction, perpendicular to the rotating core seen from high resolution observations (e.g. Sollins et al. \cite{sollins05}; Keto et al. \cite{keto88}). The outflow axis coincides with the direction of the ionised jet discovered by Keto \& Wood (\cite{keto06}).

\subsection{G16.59$-$0.06}

Our outflow maps (Fig. \ref{outflows}) suggest the presence of a bipolar outflow oriented in the NE-SW direction, with a compact blue lobe, and a more extended, less collimated red lobe.

Detailed studies of this region have been performed by several authors. In particular, Furuya et al. (\cite{furuya08}) obtained high-angular resolution \element[][12]CO(1$-$0) maps with the PdBI, which reveal a SE-NW molecular outflow. Beuther et al. (\cite{beut06}), who mapped G16.59 in the \element[][12]CO(2$-$1) line both with IRAM 30-m and with the SMA, resolve two perpendicular molecular outflows by combining their single-dish and interferometric images: one flow is oriented along the same direction reported by Furuya, and the other along the NE-SW direction. Our observations are therefore consistent with the results obtained by Beuther et al. (\cite{beut06}).

\subsection{G35.20$-$0.74}

The presence of a bipolar molecular outflow in this clump is strongly suggested by the elongated structure of the blue- and red-shifted \13co emission (Fig. \ref{outflows}). Although the blue and red outflow lobes overlap each other, an outflow axis can clearly be defined along the SW-NE direction, and the overlap could be due to an outflow inclination almost perpendicular to the line of sight. This bipolar outflow has been previously observed in CO and SiO by Gibb et al. (\cite{gibb03}). The same authors also present interferometric radio continuum maps showing a radio-jet oriented in the north-south direction, which is also detected in the mid- and near-IR by De Buizer (\cite{deb06}) and Fuller et al. (\cite{fuller01}), respectively. Gibb et al. (\cite{gibb03}) argue that the region contains up to four outflows, and consider this interpretation more plausible than precession of the jet to explain such different jet/outflow orientations.

The \c18o half power contour (white solid line in Figure \ref{outflows}) has an elongated shape which is approximately perpendicular to the \element[][13]CO outflow axis. In addition, a NW-SE velocity gradient is detected within this contour (Fig. \ref{velo}), which favours the interpretation of a flattened structure rotating perpendicularly to the outflow axis. Evidence of a large-scale disk-like structure was already presented by Little et al. (\cite{little85}), who performed \AMM\ observations at various angular resolutions and detected a flattened molecular structure with the same orientation and size as our \c18o velocity gradient.

Our observations towards G35.20 are therefore consistent with results found by other authors, and are compatible with a rotating structure seen almost edge-on and perpendicular to a bipolar molecular outflow lying close to the plane of the sky. 

\subsection{G43.89$-$0.78}

To our knowledge, no detailed studies on the kinematics of this high-mass SFR have been made in the past. From the wide wings in the \13co line (Fig. \ref{spec2}) and the spatial distribution of the blue and red lobes in Fig. \ref{outflows}, we conclude that a high-mass molecular outflow is present in this clump (see also Tables \ref{tot1} and \ref{tot2}).

Both the \c18o peak position and the centre of the molecular outflow are offset with respect to the position of the associated UC \HII\ region (Wood \& Churchwell \cite{wood89}) by about 10\arcsec, or $\sim$~0.2~pc at a distance of 4.2~kpc. This suggests that the outflow might not be driven by the massive YSO within the UC \HII\ region but by a different, presumably high-mass, young (proto)star embedded in the C$^{18}$O core.

From the \c18o first moment, a velocity gradient can be seen which is not as clear as in G10.62 or G35.20, but for which a southwest (blue) to northeast (red) direction can be assigned. This velocity gradient is neither aligned with nor perpendicular to the outflow axis (Fig. \ref{velo}), so its interpretation is unclear. Further observations at high-angular resolution would be helpful to understand the nature of the $^{13}$CO and C$^{18}$O emission.

\subsection{G75.78$+$0.34}

Shepherd et al. (\cite{shepherd97}) conducted $^{12}$CO and SiO high-resolution observations towards this source with BIMA, and the resulting maps resolve three CO bipolar molecular outflows and one possible SiO outflow within an area of approximately $1\arcmin \times 1\arcmin$ (roughly the area of the corresponding map in Figure \ref{outflows}). The axis of the reported CO outflows have a position angle (P.A.) ranging between north-south and east-west. Thus, our outflow map (Fig. \ref{outflows}), which shows no clear spatial separation between red and blue lobes, is probably englobing all three outflows detected by Shepherd et al. This is yet one more example in which the usefulness of complementary interferometric observations is evident.

More enigmatic is the presence of a \c18o velocity gradient in this clump (Figure \ref{velo}). Its west-east orientation is reversed with respect to two of the CO outflows discovered by Shepherd et al. (1997), while it is perpendicular to the third north-south outflow reported by the same authors. This would rule out the possibility that the velocity gradient is tracing an outflow, leaving us with the options of either rotation of the whole C$^{18}$O clump or multiple components within it (or both). The former would imply a dynamical mass of 23~M$_{\sun}$, too small to account for all the sources that, according to Shepherd et al. (1997), lie inside this clump. Nevertheless, we recall that this is a lower limit. Therefore the possibility of rotation is as plausible as that of source multiplicity, and the discrimination between them should perhaps be left to further high-spatial resolution images of either C$^{18}$O or high density tracers such as CH$_{3}$CN.

\end{appendix}

\end{document}